\documentclass[aps,pra,10pt,twocolumn,amsmath,amssymb,floatfix]{revtex4-1}
\usepackage{graphicx}
\usepackage{dcolumn}
\usepackage{bm}
\usepackage{amsfonts}
\usepackage{graphicx}
\usepackage{dcolumn}
\usepackage{bm}
\usepackage{hyperref}
\usepackage{amsmath}
\usepackage{amssymb}
\usepackage{verbatim}
\usepackage{epstopdf}

\usepackage[makeroom]{cancel}

\begin{document}
\title{Semiclassical Boltzmann transport theory of few-layer black phosphorus \\
in various phases}
\author{Sanghyun Park}
\author{Seungchan Woo}
\author{Hongki Min}
\email{hmin@snu.ac.kr}
\affiliation{Department of Physics and Astronomy, Seoul National University, Seoul 08826, Korea}

\date{\today}

\begin{abstract}
Black phosphorus (BP), a two-dimensional (2D) van der Waals layered material composed of phosphorus atoms, has been one of the most actively studied 2D materials in recent years due to its tunable energy band gap (tunable even to a negative value) and its highly anisotropic electronic structure. Depending on the sign of the band gap tuning parameter, few-layer BP can be in a gapped insulator phase, gapless Dirac semimetal phase, or gapless semi-Dirac transition point between the two phases.
Using the fully anisotropic multiband Boltzmann transport theory, we systematically study the dc conductivity of few-layer BP as a function of the carrier density and temperature by varying the band gap tuning parameter, and determine the characteristic density and temperature dependence corresponding to each phase.
\end{abstract}

\maketitle

\section{Introduction}
Since the discovery of graphene \cite{CastroNeto2009, DasSarma2011}, which is a carbon allotrope of two-dimensional (2D) honeycomb lattice, 2D materials have been one of the most active research areas in condensed matter physics. 
Black phosphorus (BP) is a 2D material with van der Waals layered structure composed of phosphorus atoms, and it has recently attracted considerable attention \cite{Rui2017, Chaves2017}. As a layered semiconductor in its natural form, BP has a tunable band gap, and manipulation of its band gap through various methods has been validated by multiple theoretical and experimental reports \cite{Roldan2017}. Notable examples of the band gap tuning include thickness change \cite{Yang2015, Cakir2015}, strain control \cite{Quereda2016}, pressure \cite{Xiang2015}, electronic gating \cite{Deng2017, Liu2017, Li2018}, and chemical doping \cite{Kim2015}. Some of the band gap manipulation methods \cite{Xiang2015, Kim2015} demonstrated that the band gap can be tuned to zero, showing the semi-Dirac state with a combination of linear and quadratic dispersions \cite{Banerjee2009}, which is also predicted in $\rm{TiO}_2/\rm{VO}_2$ heterostructures \cite{Pardo2009, Pardo2010}. Furthermore, the band gap can be inverted, leading to the Dirac semimetal phase \cite{Kim2017, Ehlen2018, DiPietro2018}.


Due to its anisotropic electronic band structure, BP shows many peculiar transport properties such as large in-plane anisotropic transport \cite{Xia2014, Mishchenko2015}. The effects of temperature \cite{Li2017, Illarionov2016, Deng2017}, the number of layers \cite{Deng2017}, and substrate \cite{Li2017} on the anisotropic transport properties of BP have been studied experimentally. Furthermore, the transport properties of BP have been studied theoretically \cite{Qiao2014, Liu2016, Liu2017a, Han2017, Zare2017, Adroguer2016}, demonstrating its anisotropic nature in energy- and temperature-dependent transport. 
However, there has been no systematic study on the anisotropic transport of BP in each phase, fully considering the anisotropy of the system and the interband scattering. 
In this study, we theoretically investigate the transport properties of BP in the gapped insulator phase, gapless semi-Dirac transition point, and Dirac semimetal phase. Using the semiclassical Boltzmann transport theory generalized to anisotropic multiband systems, we calculate the dc conductivity as a function of the carrier density and temperature for each phase. We determine that each phase shows the characteristic density and temperature dependence, which can be used as a transport signature of BP in different phases.

The rest of this paper is organized as follows. In Sec.~\ref{sec:theory}, we describe our model Hamiltonian and develop the Boltzmann transport theory in anisotropic multiband systems. In Sec.~\ref{sec:density_dep}, we present the dc conductivity of BP in each phase as a function of density at zero temperature. In Sec.~\ref{sec:temperature_dep}, we provide the temperature dependence of dc conductivity at a fixed density. We conclude our paper in Sec.~\ref{sec:discussion} with discussions on the dominant scattering source, the effect of potential fluctuations at low densities, and the effect of the parabolic term omitted in the current model. 

\section{Methods}
\label{sec:theory}
\subsection{Model}
\label{subsec:model}
By expanding the tight-binding lattice model of few-layer BP \cite{Chaves2017, Rudenko2014, Sousa2017}, the corresponding low-energy effective Hamiltonian can be obtained as \cite{Baik2015, Doh2017, Montambaux2009a, Montambaux2009b, deGail2012}
\begin{equation}\label{eq:BP}
H=\left(\frac{\hbar^2 k_{x}^2}{2m^{*}}+\frac{\varepsilon_{\rm g}}{2} \right) \sigma_{x} + \hbar v_0 k_{y} \sigma_{y} ,
\end{equation}
where $m^{*}$ is the effective mass along the zigzag ($x$) direction, $v_0$ is the band velocity along the armchair ($y$) direction, $\varepsilon_{\rm g}$ is the size of the band gap (which will be used as a tuning parameter), and $\sigma_x$ and $\sigma_y$ are the Pauli matrices. The eigenenergies of the Hamiltonian are given by $\varepsilon_{\pm}=\pm\sqrt{\left(\frac{\hbar^2 k_{x}^2}{2m^{*}}+\frac{\varepsilon_{\rm g}}{2} \right)^2 +  \hbar^2 v_0^2 k_{y}^2}$; thus, the Hamiltonian $H$ has a direct band gap for $\varepsilon_{\rm g}>0$, a semi-Dirac band touching point at $(k_x,k_y)=(0,0)$ for $\varepsilon_{\rm g}=0$, or two Dirac points at $(k_x,k_y)=(\pm \sqrt{\frac{m^{*}|\varepsilon_{\rm g}|}{\hbar^2}},0)$ for $\varepsilon_{\rm g}<0$. 
The characteristic energy scales along the zigzag and armchair directions are given by $\varepsilon_0={\hbar^2 k_0^2\over 2m^{*}}$ and $\hbar v_0 k_0$, respectively, where $k_0=a^{-1}$ and $a$ is the lattice constant. We introduce the dimensionless parameters $\Delta={\varepsilon_{\rm g}\over 2\varepsilon_0}$ and $c={\hbar v_0 k_0\over \varepsilon_0}$, which represent a gap tuning parameter and the ratio of the characteristic energy scales along the zigzag and armchair directions, respectively.
Throughout the paper, we use $c=1$ and the spin degeneracy $g=2$ for the calculation. We will discuss the effect of higher-order terms omitted in Eq.~(\ref{eq:BP}) in Sec.~\ref{sec:discussion}. 


\begin{figure}[htb]
\includegraphics[width=\linewidth]{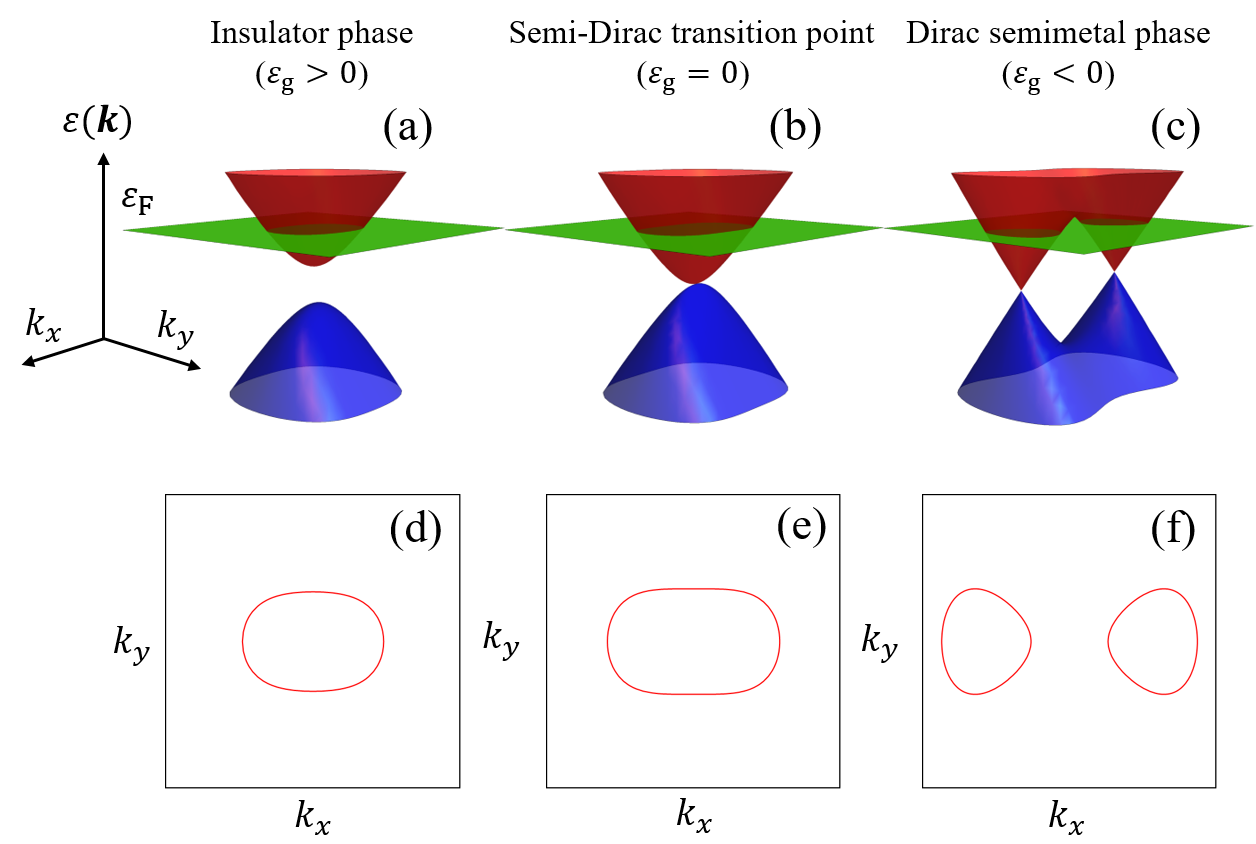}
\caption{
(a)-(c) Energy dispersions and (d)-(f) the corresponding Fermi surfaces of few-layer BP for the (a), (c) insulator phase, (b), (e) semi-Dirac transition point, and (c), (f) Dirac semimetal phase.
}
\label{fig:dispersion}
\end{figure}

Figure~\ref{fig:dispersion} shows the energy dispersion and the corresponding Fermi surface of few-layer BP in each phase. Initially, few-layer BP without band gap tuning is in the gapped insulator phase, as shown in Fig.~\ref{fig:dispersion}(a). As the band gap $\varepsilon_{\rm g}$ decreases (for example, upon applying a perpendicular electric field), eventually it vanishes and the system is described by the semi-Dirac Hamiltonian in Eq.~(\ref{eq:BP}) with $\varepsilon_{\rm g}=0$, as  shown in Fig.~\ref{fig:dispersion}(b). 
If the band gap decreases even further and becomes negative ($\varepsilon_{\rm g}<0$), band inversion occurs, 
which has been achieved experimentally using surface doping \cite{Kim2017, Ehlen2018} and external pressure \cite{DiPietro2018} 

In the gapped insulator phase, the inherent anisotropy of the system is less evident and the system at low densities resembles typical semiconductors with a different effective mass in each direction. At the semi-Dirac transition point, the energy dispersion becomes linear (quadratic) along the armchair (zigzag) direction, 
as shown in Fig.~\ref{fig:dispersion}(e). At the Dirac semimetal phase, the anisotropy in the energy dispersion becomes more pronounced and the Fermi surface vastly changes its shape depending on the value of the Fermi energy $\varepsilon_{\rm F}$. For $\varepsilon_{\rm F}<\varepsilon_{\rm g}/2$, the Fermi surface becomes two distinct lines, as shown in Fig.~\ref{fig:dispersion}(f), whereas for $\varepsilon_{\rm F}>\varepsilon_{\rm g}/2$, the two Fermi surfaces become joined completely, forming a closed line. At $\varepsilon_{\rm F}=\varepsilon_{\rm g}/2$, a van Hove singularity occurs in the density of states (DOS), as explained below.

\begin{figure}[htb]
\includegraphics[width=\linewidth]{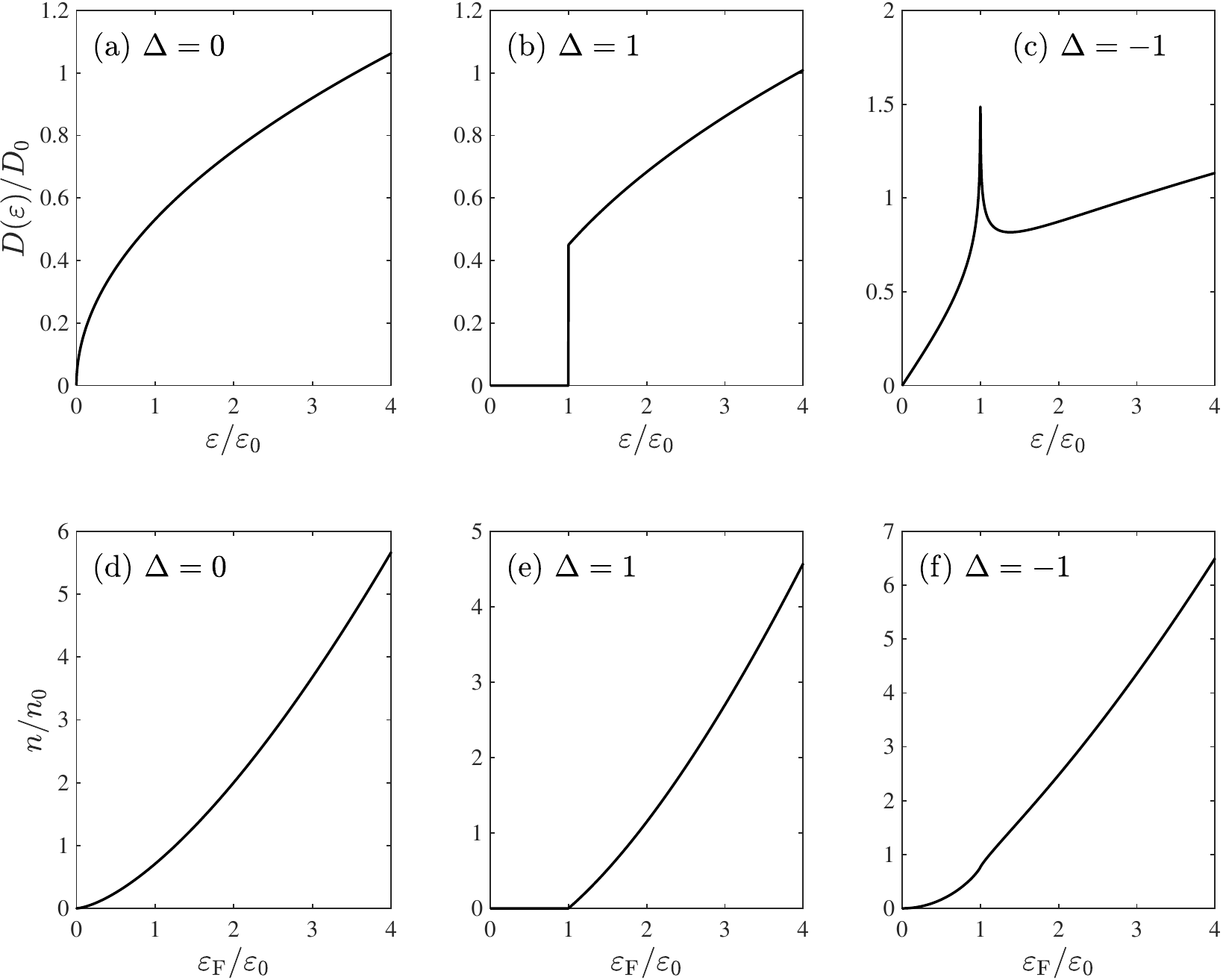}
\caption{
(a)-(c) Calculated DOS and (d)-(e) the carrier density as a function of Fermi energy for the (a), (c) insulator phase, (b), (e) semi-Dirac transition point, and (c), (f) Dirac semimetal phase. Here, $\Delta\equiv \frac{\varepsilon_{\rm g}}{2\varepsilon_0}$ is the band gap tuning parameter, and $g=2$ and $c=1$ are used for calculation. 
}
\label{fig:DOS_density}
\end{figure}

Figure~\ref{fig:DOS_density} shows the DOS and the carrier density as a function of Fermi energy for each phase. At the semi-Dirac transition point, the DOS is simply given by $D(\varepsilon)\sim\varepsilon^{1/2}$ [Fig.~\ref{fig:DOS_density}(a)], and the carrier density (which is an energy integral of the DOS up to $\varepsilon_{\rm F}$) is given by $n\sim \varepsilon_{\rm F}^{3/2}$ [Fig.~\ref{fig:DOS_density}(d)]. (See Appendix \ref{sec:eigenstate_dos} for the detailed derivations of the DOS and the carrier density.) 
In the gapped insulator phase, both DOS at $\varepsilon_{\rm F}$ and carrier density vanish for $\varepsilon_{\rm F}<\varepsilon_{\rm g}/2$, whereas for $\varepsilon_{\rm F}>\varepsilon_{\rm g}/2$, they follow those of the semi-Dirac transition point as $\varepsilon_{\rm F}$ increases [Figs.~\ref{fig:DOS_density}(b) and \ref{fig:DOS_density}(e)].
In the Dirac semimetal phase, when $\varepsilon_{\rm F}$ is very small, the system resembles a typical 2D Dirac semimetal such as graphene; thus, $D(\varepsilon)\sim \varepsilon^{1}$. As $\varepsilon_{\rm F}$ increases and approaches $\varepsilon_{\rm g}/2$ near the top of the inverted band, the band dispersion effectively becomes hyperbolic paraboloid with a different sign in each direction in momentum space. Subsequently, a van Hove singularity occurs in the DOS, diverging logarithmically with $D(\varepsilon)\sim -\log(|\Delta|-\varepsilon)^{-1}$ \cite{Marder2010}. If $\varepsilon_{\rm F}$ increases further, the DOS and the carrier density follow those of the semi-Dirac transition point with a discontinuous energy derivative in the DOS at the van Hove singularity [Figs.~\ref{fig:DOS_density}(c) and \ref{fig:DOS_density}(f)].

Notably, as the energy dispersion and the Fermi surface are anisotropic, and the Fermi energy can cross multiple bands, we cannot naively use the conventional Boltzmann transport theory assuming an isotropic single-band system. Thus, the anisotropic multiband Boltzmann transport theory is necessary to calculate the dc conductivity of such systems, as explained in Sec.~\ref{subsec:Boltzmann_aniso_multiband}.

\subsection{Boltzmann transport theory in anisotropic multiband systems}
\label{subsec:Boltzmann_aniso_multiband}
We use semiclassical Boltzmann transport theory to calculate the density and temperature dependence of the dc conductivity of few-layer BP in each phase in the presence of impurities, assuming elastic scattering (see Sec.~\ref{sec:discussion} for the limitation of the current approach).
In the Boltzmann transport theory, electron states are described by the non-equilibrium distribution function $f=f(\bm{r},\bm{k};t)$. Its time rate of change is balanced out by the collision term, which represents the total scattering probability per unit time, i.e., $\frac{df}{dt}=\left(\frac{df}{dt}\right)_{\rm coll}$.

We assume a spatially homogeneous system without explicit time dependence in the distribution function, i.e., $f=f_{\bm{k}}$.
Thus, the time derivative of the distribution function is given by $\frac{df}{dt}=\dot{\bm{k}}\cdot\frac{\partial f_{\bm k}}{\partial \bm{k}}$, whereas the collision term is given by
\begin{equation}\label{eq:coll_total}
\left(\frac{df}{dt}\right)_{\rm coll}=-\int {\frac{d^d k'}{(2\pi)^d}} W_{\bm{k}\bm{k}'}(f_{\bm{k}}-f_{\bm{k}'}),
\end{equation} 
where $W_{\bm{k}\bm{k}'}=\frac{2\pi}{\hbar} n_{\rm imp} |V_{\bm{k}\bm{k}'}|^2 \delta(\varepsilon_{\bm{k}}-\varepsilon_{\bm{k}'})$ is the transition rate from $\bm{k}$ to $\bm{k}'$  for an elastic scattering with the impurity potential $V_{\bm{k}\bm{k}'}$ and the impurity density $n_{\rm imp}$.
In the presence of a uniform electric field $\bm{E}$, $\hbar \dot{\bm{k}}=(-e)\bm{E}$, and to the leading order in $\bm{E}$, 
\begin{equation}\label{eq:diffusion}
\frac{df_{\bm{k}}}{dt} \approx (-e) \bm{E} \cdot\frac{\partial f_{\bm{k}}^{(0)}}{\hbar \partial \bm{k}} = (-e) \bm{E} \cdot \bm{v}_{\bm{k}} \frac{\partial f_{\bm{k}}^{(0)}}{\partial \varepsilon_{\bm{k}}},
\end{equation}
where $\bm{v}_{\bm{k}}=\frac{1}{\hbar}\frac{\partial \varepsilon_{\bm{k}}}{\partial\bm{k}}$ and $f^{(0)}_{\bm{k}}=f^{(0)}(\varepsilon_{\bm k})=\left[e^{\beta (\varepsilon_{\bm k}-\mu)}+ 1\right]^{-1}$ is the Fermi--Dirac distribution function at equilibrium with $\beta=\frac{1}{k_{\rm B} T}$ and the chemical potential $\mu$. 
Assume that, to the leading order in $\bm{E}$, the non-equilibrium distribution function $f_{\bm{k}}$ is given by $f_{\bm{k}}\equiv f^{(0)}(\varepsilon)+\delta f_{\bm{k}}$ at energy $\varepsilon=\varepsilon_{\bm k}$. Thus, from $\frac{df}{dt}=\left(\frac{df}{dt}\right)_{\rm coll}$, we obtain 
\begin{equation}\label{eq:boltzmann_equality_simple}
(-e) \bm{E} \cdot \bm{v}_{\bm{k}} S^{(0)}(\varepsilon)=\int {\frac{d^d k'}{(2\pi)^d}} W_{\bm{k}\bm{k}'}(\delta f_{\bm{k}}-\delta f_{\bm{k}'}),
\end{equation}
where $S^{(0)}(\varepsilon)=-\frac{\partial f^{(0)}(\varepsilon)}{\partial \varepsilon}$.
If the Fermi energy crosses multiple energy bands, Eq.~(\ref{eq:boltzmann_equality_simple}) is generalized to \cite{Siggia1970,Woo2017}
\begin{equation}
(-e)\bm{E}\cdot\bm{v}_{\bm{k}\alpha}S^{(0)}(\varepsilon)=\sum_{\alpha'} \int \frac{d^d k'}{(2\pi)^d} W^{\alpha \alpha'}_{\bm{k}\bm{k}'}\left(\delta f_{\bm{k}}^{\alpha}-\delta f_{\bm{k}'}^{\alpha'}\right),
\end{equation}
where $\alpha$ and $\alpha'$ are band indices. 

We parameterize $\delta f^{\alpha}_{\bm{k}}$ in the following form \cite{Schliemann2003, Vyborny2009,Park2017}:
\begin{equation}
\delta f_{\bm{k}}^{\alpha}=(-e)\left(\sum_{i=1}^{d} E^{(i)} v_{\bm{k}\alpha}^{(i)} \tau_{\bm{k}\alpha}^{(i)} \right)S^{(0)}(\varepsilon),
\end{equation}
where $E^{(i)}$, $v_{\bm{k}\alpha}^{(i)}$, and $\tau_{\bm{k}\alpha}^{(i)}$ are the electric field, velocity, and relaxation time, respectively, along the $i$th direction for each band.
After matching each coefficient in $E^{(i)}$, we obtain the following integral equation for the relaxation time:
\begin{equation}
\label{eq:relaxation_time_anisotropic}
1=\sum_{\alpha'}\int \frac{d^d k'}{(2\pi)^d} W^{\alpha\alpha'}_{\bm{k}\bm{k}'}\left(\tau_{\bm{k}\alpha}^{(i)}-\frac{v_{\bm{k}'\alpha'}^{(i)}}{v_{\bm{k}\alpha}^{(i)} }\tau_{\bm{k}'\alpha'}^{(i)}\right).
\end{equation}
This is a coupled integral equation relating the relaxation times at different angles in different bands, which correctly considers the anisotropy and multiple bands of the system. Note that, for an isotropic single-band system [$\tau_{\bm{k}\alpha}^{(i)}=\tau(\varepsilon$) for a given energy $\varepsilon=\varepsilon_{\bm{k} \alpha}$], Eq.~(\ref{eq:relaxation_time_anisotropic}) is reduced to the well-known expression for the relaxation time given by \cite{Ashcroft1976}
\begin{equation}
\label{eq:relaxation_time_isotropic}
{\frac{1}{\tau_{\bm k}}}=\int {\frac{d^d k'}{(2\pi)^d}} W_{\bm{k}\bm{k}'} (1-\cos\theta_{\bm{k}\bm{k}'}).
\end{equation}

The current density $\bm{J}$ induced by an electric field $\bm{E}$ is thus given by
\begin{equation}
J^{(i)}=g\sum_{\alpha}\int \frac{d^d k}{(2\pi)^d} (-e) v_{\bm{k}\alpha}^{(i)} \delta f_{\bm{k}\alpha}\equiv \sum_j \sigma_{ij}E^{(j)},
\end{equation}
where $\sigma_{ij}$ is the conductivity tensor given by
\begin{equation}
\label{eq:conductivity_tensor}
\sigma_{ij}=g e^2\sum_{\alpha}\int \frac{d^d k}{(2\pi)^d} S^{(0)}(\varepsilon) v_{\bm{k}\alpha}^{(i)} v_{\bm{k}\alpha}^{(j)}\tau_{\bm{k}\alpha}^{(j)}.
\end{equation}
We find that the Hall conductivity ($i\neq j$) vanishes, thus we consider only the diagonal part of the dc conductivity ($i=j$).


\section{Density dependence of dc conductivity}
\label{sec:density_dep}

Using the anisotropic multiband Boltzmann transport theory developed in Sec.~\ref{subsec:Boltzmann_aniso_multiband}, we calculate the dc conductivity of few-layer BP as a function of the carrier density or Fermi energy at zero temperature for each phase: the semi-Dirac transition point ($\Delta=0$), gapped insulator phase ($\Delta>0$), and Dirac semimetal phase ($\Delta<0$), all of which can be expressed by Eq.~(\ref{eq:BP}). 

As for the impurity potential, we consider two types of impurity scattering: short-range impurities and long-range Coulomb impurities (or charged impurities). Short-range impurities originate from lattice defects, vacancies, dislocations, etc., and their potential form is given by a constant in momentum space, $V_{\bm{k}\bm{k}'}=V_{\rm short}$, as they are approximately represented by the delta function in real space. For charged impurities distributed randomly in the background, the impurity potential is given by $V_{\bm{k}\bm{k}'}=\frac{2\pi e^2}{\epsilon(\bm{q})|\bm{q}|}$ in 2D, where $\epsilon(\bm{q})$ is the dielectric function for $\bm{q}=\bm{k}-\bm{k}'$. Within the Thomas--Fermi approximation, $\epsilon(\bm{q})$ can be approximated as $\epsilon(\bm{q})\approx \kappa\left(1+q_{\rm TF}/|\bm{q}| \right)$, where $\kappa$ is the background dielectric constant, $q_{\rm TF}=\frac{2\pi e^2}{\kappa} D(\varepsilon_{\rm F})$ is the Thomas--Fermi wave vector, and $D(\varepsilon_{\rm F})$ is the total DOS at the Fermi energy $\varepsilon_{\rm F}$ (including all the contributions from the bands crossing $\varepsilon_{\rm F}$ and the spin degeneracy). The interaction strength for charged impurities can be characterized by an effective fine structure constant $\alpha_0=\frac{e^2}{\kappa \hbar v_0}$. Note that $q_{\rm TF}\propto g\alpha_0$. 
Thus, the screening strength for Coulomb impurities is also characterized by $\alpha_0$.

\subsection{Semi-Dirac transition point}

\begin{figure}[htb]
\includegraphics[width=\linewidth]{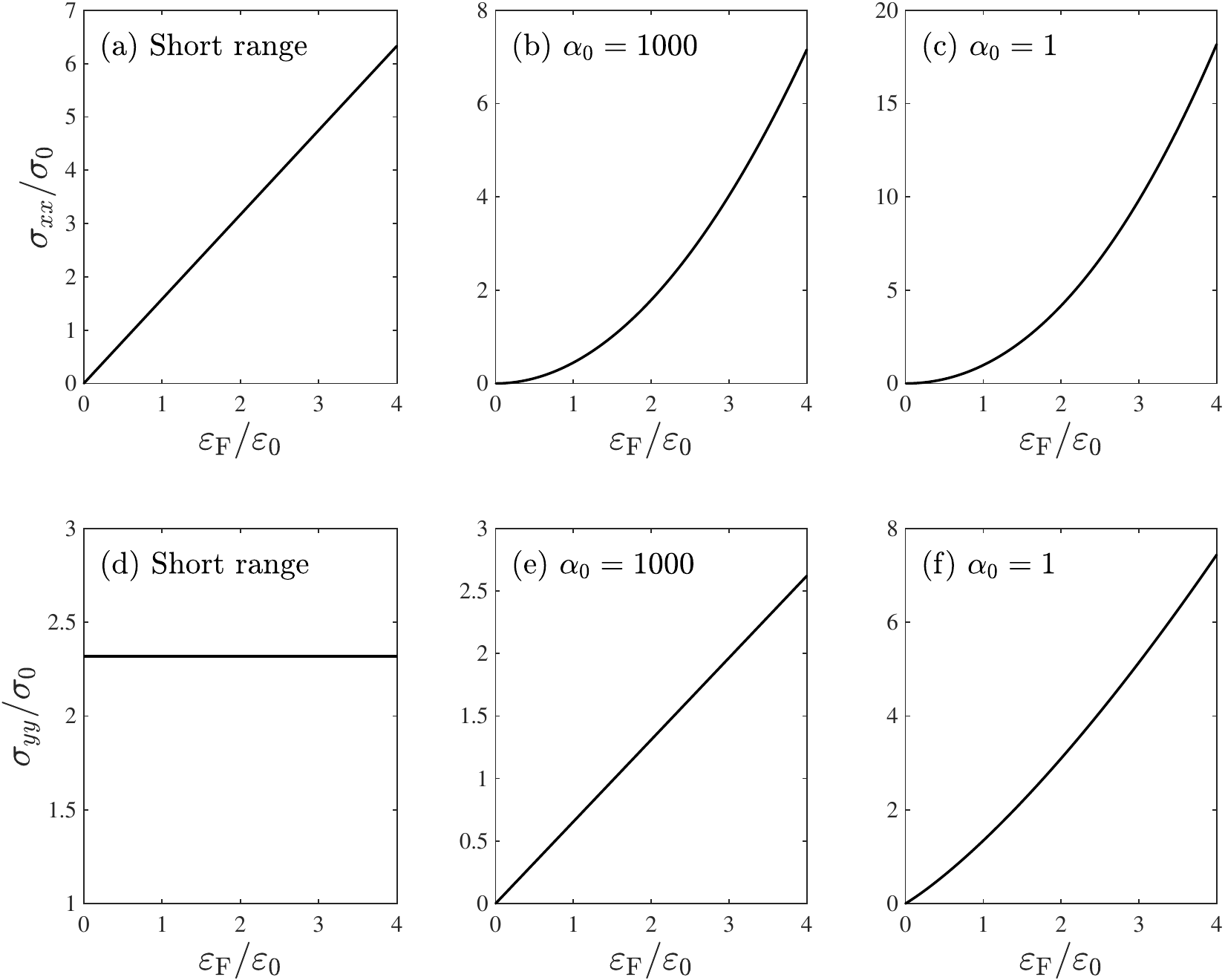}
\caption{
Calculated dc conductivities (a)-(c) $\sigma_{xx}$ and (d)-(f) $\sigma_{yy}$ as a function of Fermi energy at the semi-Dirac transition point ($\Delta=0$) for (a), (d) short-range impurities,  (b), (d) charged impurities with $\alpha_0=1000$, and (c), (f) charged impurities with $\alpha_0=1$. Here, $\sigma_0= \frac{g e^2 k_0^2 c^2}{2\pi  \hbar n_{\rm imp}}$.
}
\label{fig:density_semidirac}
\end{figure}

First, let us consider the semi-Dirac transition point ($\Delta=0$). Figure \ref{fig:density_semidirac} shows the Fermi energy dependence of dc conductivity at the semi-Dirac transition point.
The characteristic density or Fermi energy dependence of the dc conductivity can be understood as follows. 
From Eq.~(\ref{eq:conductivity_tensor}) with $\tau_{\rm F}^{(i)}\sim D^{-1}(\varepsilon_{\rm F})/V_{\rm F}^{2}$, we expect $\sigma_{ii}\sim D(\varepsilon_{\rm F}) [v_{\rm F}^{(i)}]^2 \tau_{\rm F}^{(i)}\sim [v_{\rm F}^{(i)}]^2/V_{\rm F}^2$, where $\tau_{\rm F}^{(i)}$ and $v_{\rm F}^{(i)}$ are the relaxation time and velocity, respectively, at the Fermi energy along the $i$th direction, and $V_{\rm F}^2$ is the angle-averaged squared impurity potential at the Fermi energy. 
At the semi-Dirac transition point, $D(\varepsilon_{\rm F})\sim \varepsilon_{\rm F}^{1/2}$, and the Fermi velocity in each direction is given by $v_{\rm F}^{(x)} \sim \varepsilon_{\rm F}^{1/2}$ and $v_{\rm F}^{(y)} \sim \varepsilon_{\rm F}^{0}$, from which we can deduce the power-law behavior of the dc conductivity. (See Appendix \ref{sec:eigenstate_dos} for the detailed derivations of the power-law dependences.) 

For short-range impurities, $V_{\rm F}$ is a constant independent of density; in this case, we obtain
\begin{subequations}\label{eq:density_dependence_short_semidirac}
\begin{eqnarray}
\sigma_{xx}&\sim& \varepsilon_{\rm F}^{}\sim n^\frac{2}{3}, \\
\sigma_{yy}&\sim& \varepsilon_{\rm F}^{0} \sim n^0.
\end{eqnarray}
\end{subequations}
For charged impurities, in the strong screening limit ($g\alpha_0\gg 1$), $V_{\rm F}\sim q_{\rm TF}^{-1}\sim D^{-1}(\varepsilon_{\rm F})\sim \varepsilon_{\rm F}^{-\frac{1}{2}}$; thus, we obtain
\begin{subequations}\label{eq:density_dependence_charged_strong_screening_semidirac}
\begin{eqnarray}
\sigma_{xx}&\sim& \varepsilon_{\rm F}^{2} \sim n^\frac{4}{3}, \\
\sigma_{yy}&\sim& \varepsilon_{\rm F}^{}\sim n^\frac{2}{3}.
\end{eqnarray}
\end{subequations}
At general screening strength, the power-law behavior is determined by the competition between the screening wave vector and the momentum transfer. 
We present the numerically calculated power-law behavior for the semi-Dirac transition point and for the other phases in Fig.~\ref{fig:density_powever_evolution}.

\subsection{Insulator phase}

\begin{figure}[htb]
\includegraphics[width=\linewidth]{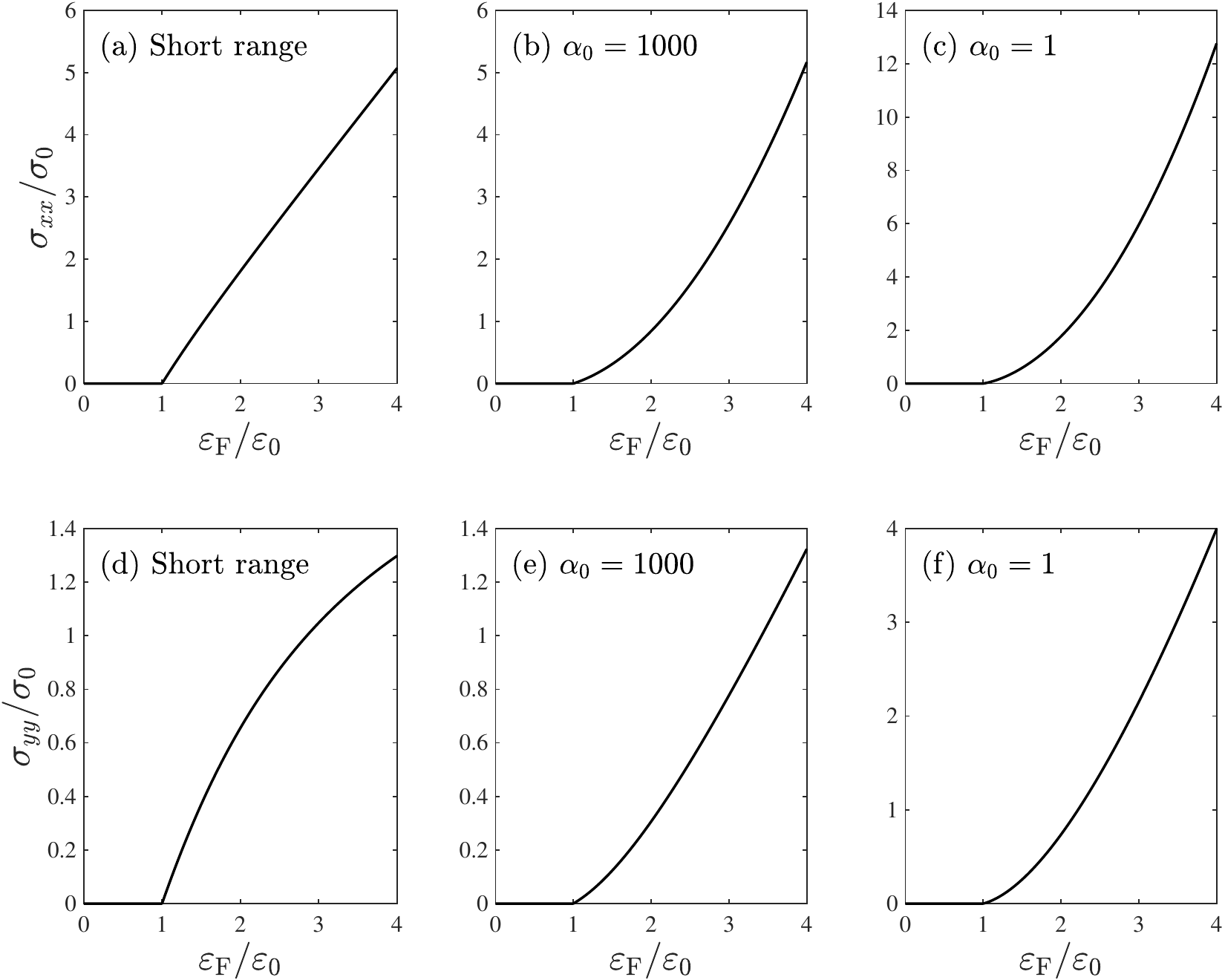}
\caption{
Calculated dc conductivities (a)-(c) $\sigma_{xx}$ and (d)-(f) $\sigma_{yy}$ as a function of Fermi energy in the insulator phase with $\Delta=1$ for (a), (d) short-range impurities,  (b), (d) charged impurities with $\alpha_0=1000$, and (c), (f) charged impurities with $\alpha_0=1$. 
}
\label{fig:density_semicon}
\end{figure}

Figure~\ref{fig:density_semicon} shows the Fermi energy dependence of the dc conductivity in the insulator phase ($\Delta>0$). In the insulator phase, the power-law dependence of the dc conductivity at low densities becomes similar to that of 2D electron gas (2DEG) with a different effective mass in each direction. (See Appendix \ref{subsec:eff_semicon} for detailed derivations.) 

For short-range impurities, the power-law dependence of the dc conductivity at low densities is given by
\begin{subequations}\label{eq:density_dependence_short_semicon}
\begin{eqnarray}
\sigma_{xx}&\sim& \varepsilon_{\rm F}^{}, \\
\sigma_{yy}&\sim& \varepsilon_{\rm F}^{}.
\end{eqnarray}
\end{subequations}
For charged impurities, in the strong screening limit, at low densities, we obtain
\begin{subequations}\label{eq:density_dependence_charged_strong_screening_semicon}
\begin{eqnarray}
\sigma_{xx}&\sim& \varepsilon_{\rm F}^{}, \\
\sigma_{yy}&\sim& \varepsilon_{\rm F}^{}.
\end{eqnarray}
\end{subequations}

Note that, as the Fermi energy or the carrier density increases, the power-law dependence becomes similar to that of the semi-Dirac transition point.

\subsection{Dirac semimetal phase}

\begin{figure}[htb]
\includegraphics[width=\linewidth]{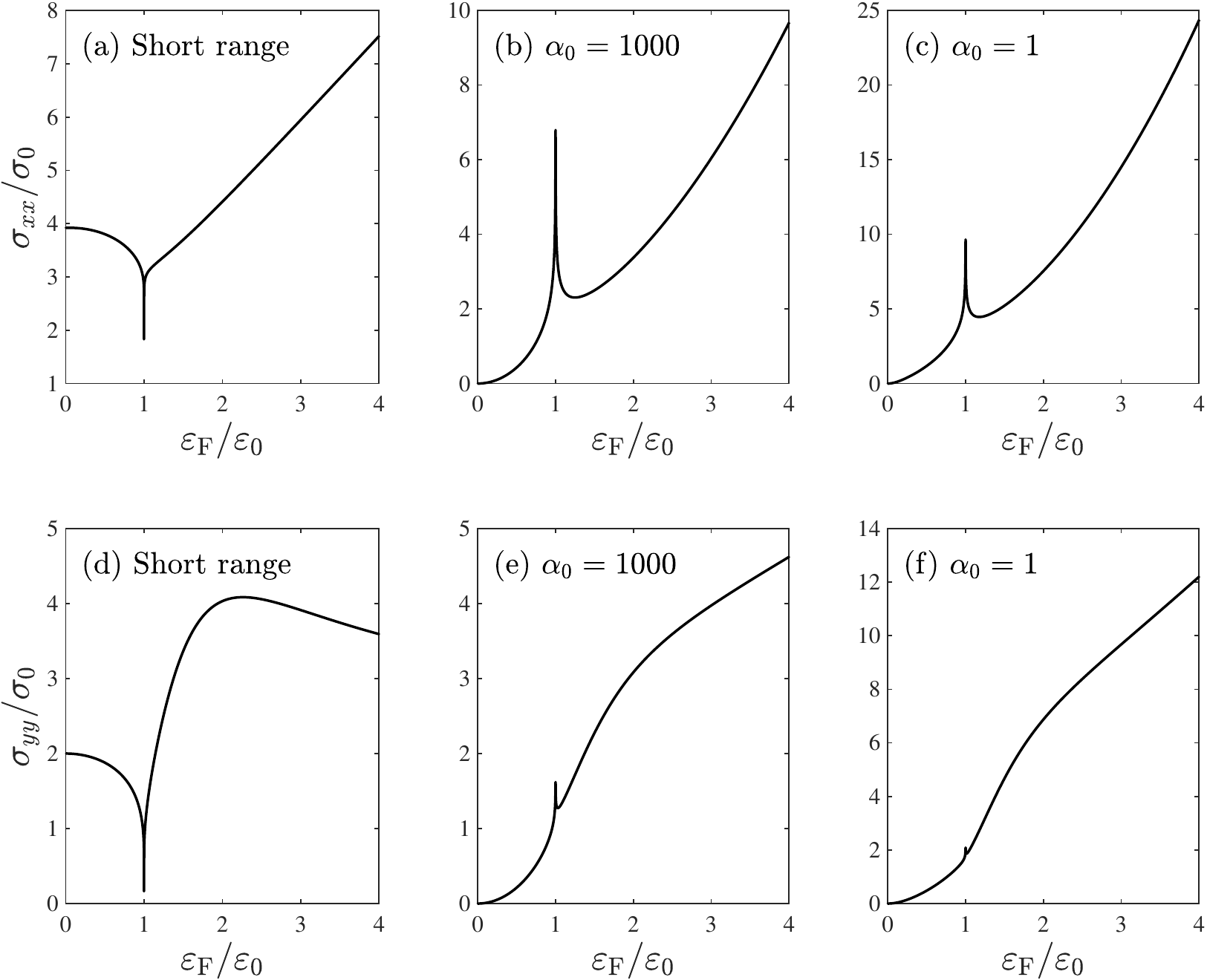}
\caption{
Calculated dc conductivities (a)-(c) $\sigma_{xx}$ and (d)-(f) $\sigma_{yy}$ as a function of Fermi energy in the Dirac semimetal phase with $\Delta=-1$ for (a), (d) short-range impurities,  (b), (d) charged impurities with $\alpha_0=1000$, and (c), (f) charged impurities with $\alpha_0=1$. 
}
\label{fig:density_dirac}
\end{figure}

Figure~\ref{fig:density_dirac}	shows the Fermi energy dependence of the dc conductivity in the Dirac semimetal phase ($\Delta<0$). In the Dirac semimetal phase, the power-law dependence of the dc conductivity at low densities becomes similar to that of graphene but with a different Fermi velocity in each direction. (See Appendix \ref{subsec:eff_dirac} for detailed derivations.) 

For short-range impurities, the power-law dependence of the dc conductivity at low densities is given by
\begin{subequations}\label{eq:density_dependence_short_Dirac}
\begin{eqnarray}
\sigma_{xx}&\sim& \varepsilon_{\rm F}^{0}, \\
\sigma_{yy}&\sim& \varepsilon_{\rm F}^{0}.
\end{eqnarray}
\end{subequations}
For charged impurities, in the strong screening limit, at low densities, we obtain
\begin{subequations}\label{eq:density_dependence_charged_strong_screening_Dirac}
\begin{eqnarray}
\sigma_{xx}&\sim& \varepsilon_{\rm F}^{2}, \\
\sigma_{yy}&\sim& \varepsilon_{\rm F}^{2}.
\end{eqnarray}
\end{subequations}

Near the van Hove singularity, $\varepsilon_{\rm F}\approx \pm \varepsilon_{\rm g}/2$, the DOS diverges logarithmically \cite{Marder2010} and it dominates the overall power-law behavior of conductivity \cite{Woo2017}. Therefore, for short-range impurities, the conductivity becomes 
\begin{subequations}\label{eq:density_dependence_short_vanhove}
\begin{eqnarray}
\sigma_{xx}&\sim& \left[-\log\left(|\Delta|-\varepsilon_{\rm F}\right)\right]^{-1}, \\
\sigma_{yy}&\sim& \left[-\log\left(|\Delta|-\varepsilon_{\rm F}\right)\right]^{-1}.
\end{eqnarray}
\end{subequations}
For charged impurities, due to the dominant contribution from the diverging Thomas--Fermi wave vector $q_{\rm TF}\propto D(\varepsilon_{\rm F})$, the conductivity is largely given by the square of the DOS as follows:
\begin{subequations}\label{eq:density_dependence_charged_strong_screening_vanhove}
\begin{eqnarray}
\sigma_{xx}&\sim& \left[\log\left(|\Delta|-\varepsilon_{\rm F}\right)\right]^{2}, \\
\sigma_{yy}&\sim& \left[\log\left(|\Delta|-\varepsilon_{\rm F}\right)\right]^{2}.
\end{eqnarray}
\end{subequations}

As the Fermi energy or the carrier density increases further, the power-law dependence of the dc conductivity becomes similar to that of the semi-Dirac transition point, as in the insulator phase.

\begin{figure}[htb]
\includegraphics[width=\linewidth]{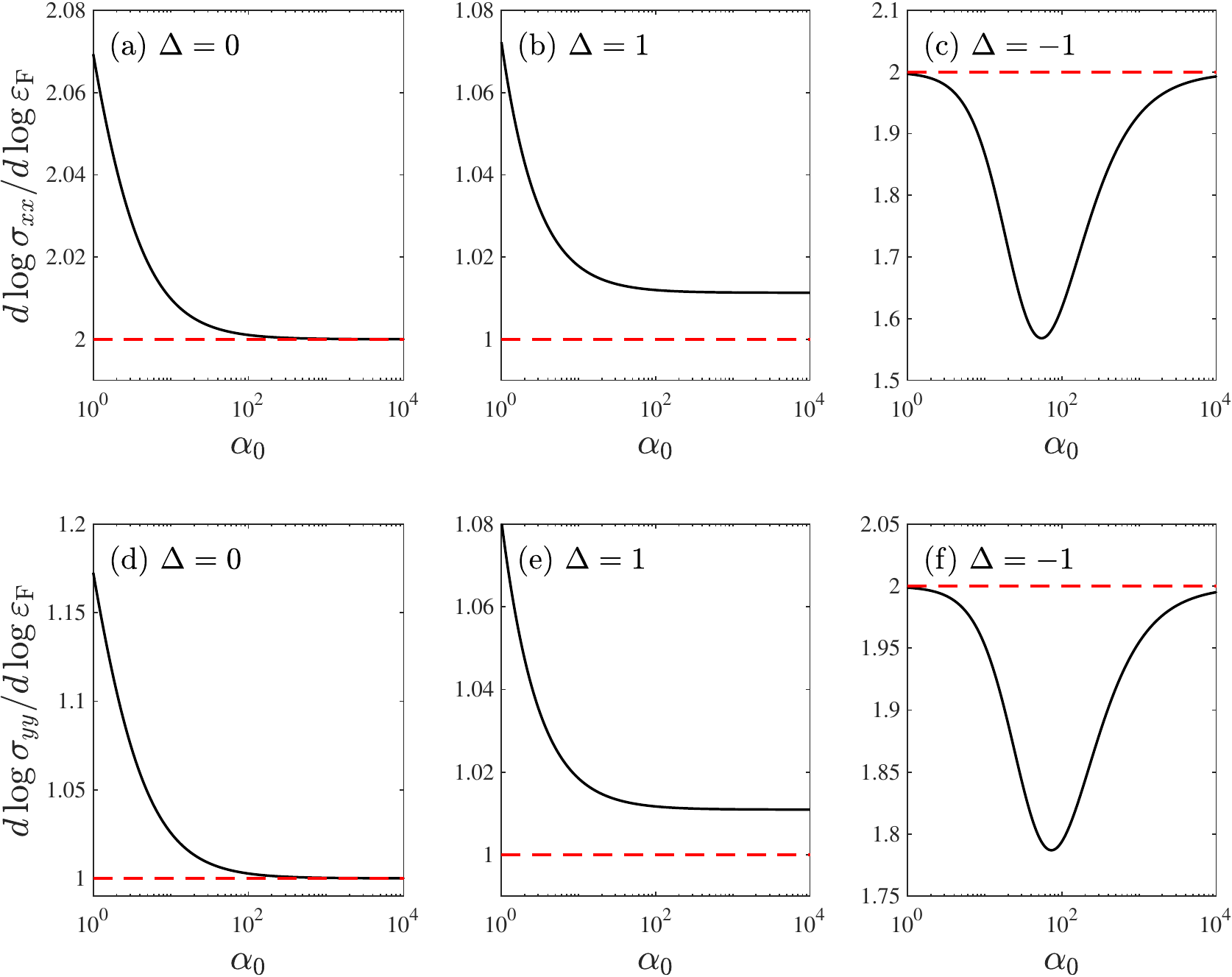}
\caption{
(a)-(c) $d\log \sigma_{xx}/d\log \varepsilon_{\rm F}$ and (d)-(f) $d\log \sigma_{yy}/d\log \varepsilon_{\rm F}$ as a function of $\alpha_0$ for charged impurities in each phase. The red dashed lines represent the Fermi energy exponents obtained in the strong screening limit. Here, $\varepsilon_{\rm F}=\varepsilon_0$ for the semi-Dirac transition point, $\varepsilon_{\rm F}=1.01\varepsilon_0$ for the gapped insulator phase, and $\varepsilon_{\rm F}=0.01\varepsilon_0$ for the Dirac phase are used for the calculation.
}
\label{fig:density_powever_evolution}
\end{figure}

Figure~\ref{fig:density_powever_evolution} shows the evolution of the Fermi-energy power law of the dc conductivity as a function of the screening strength $\alpha_0$ for each phase in the low carrier density limit. 
For the insulator phase and the semi-Dirac transition point, the Fermi-energy exponent decreases, whereas for the Dirac semimetal phase, it shows a non-monotonic behavior with a dip structure, which originates from the interband-like scattering between two distinct Fermi surfaces shown in Fig.~\ref{fig:dispersion}(f). 
As the screening strength increases, all the Fermi-energy exponents approach the corresponding power law estimated in the strong screening limit.

\section{Temperature dependence of dc conductivity}
\label{sec:temperature_dep}

We can apply the anisotropic multiband Boltzmann transport theory developed in Sec.~\ref{subsec:Boltzmann_aniso_multiband} to the dc conductivity at finite temperature. In Eq.~(\ref{eq:conductivity_tensor}), the finite temperature affects the conductivity through the Fermi distribution and the temperature-dependent screening
for the charged impurity potential. At finite temperatures, the chemical potential of the system also deviates from the Fermi energy $\varepsilon_{\rm F}$ due to the broadening of the Fermi distribution function. From the invariance of carrier density $n$ with respect to temperature $T$, we obtain the temperature dependence of the chemical potential $\mu(T)$. For charged impurities, the finite temperature Thomas--Fermi screening wave vector is given by $q_{\rm TF}(T)=\frac{2\pi e^2}{\kappa} \frac{\partial n}{ \partial \mu}$ for 2D systems. (See Appendix \ref{sec:chemical_potential_screening_wavevector} for the detailed derivation of the temperature dependence of the chemical potential and Thomas--Fermi wave vector). In this section, we calculate the dc conductivity of few-layer BP as a function of the temperature for each phase. The detailed derivation of the temperature-dependent conductivity is presented in Appendices \ref{sec:app_temperature_dep_semidirac} and \ref{sec:app_temperature_dep_effmodel}.

\subsection{Semi-Dirac transition point}
\label{temperature_dependence_semi-Dirac}
From the power-law dependence of the DOS, $D(\varepsilon)\sim\varepsilon^{1/2}$ at the semi-Dirac transition point [Fig.~\ref{fig:DOS_density}(a)], we can obtain the asymptotic behaviors of $\mu(T)$ and $q_{\rm TF}(T)$ in a relatively straightforward manner. In the low- and high-temperature limits, the chemical potential at the semi-Dirac transition point is given by 
\begin{eqnarray}\label{eq:asymptotic_mu_semidirac_main}
{\frac{\mu}{\varepsilon_{\rm F}}} &=&
\begin{cases}
1- \frac{\pi^2}{12} \left(\frac{T}{T_{\rm F}} \right)^2 & (T\ll T_{\rm F}), \\ {1\over 2\eta\left({1\over 2}\right)\Gamma\left({5\over 2}\right)}\left(\frac{T}{T_{\rm F}}\right)^{{1\over 2}} & (T\gg T_{\rm F}),
\end{cases}
\end{eqnarray}
whereas the Thomas--Fermi wave vector is given by
\begin{eqnarray}\label{eq:asymptotic_qtf_semidirac_main}
{\frac{q_{\rm{TF}}(T) }{q_{\rm{TF}}(0) }} &= &
\begin{cases}
1- \frac{\pi^2}{12} \left(\frac{T}{T_{\rm F}} \right)^2  & (T\ll T_{\rm F}), \\
2\eta\left({1\over 2}\right)\Gamma\left({3\over 2}\right)\left(\frac{T}{T_{\rm F}}\right)^{1\over 2} & (T\gg T_{\rm F}),
\end{cases}
\end{eqnarray}
where $\Gamma$ is the Gamma function and $\eta$ is the Dirichlet eta function \cite{Arfken2012}. 
In a single-band system, $q_{\rm TF}(T)$ typically decreases with the temperature at high temperatures, whereas at the semi-Dirac transition point, $q_{\rm TF}(T)$ increases with the temperature due to the thermal excitation of carriers participating in the screening.

\begin{figure}[htb]
\includegraphics[width=\linewidth]{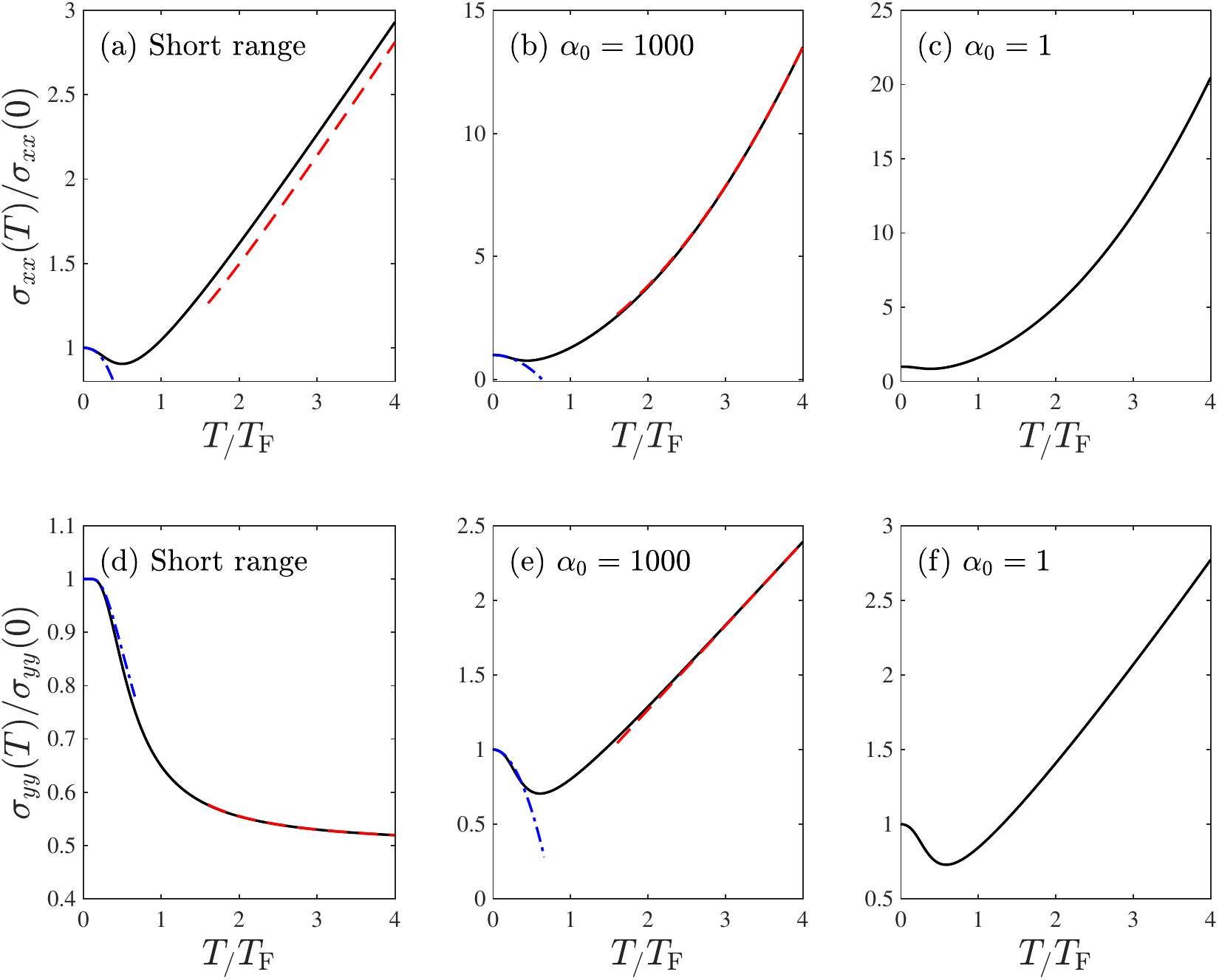}
\caption{
Calculated dc conductivities (a)-(c) $\sigma_{xx}$ and (d)-(f) $\sigma_{yy}$ as a function of the temperature at the semi-Dirac transition point ($\Delta=0$) for (a), (d) short-range impurities, (b), (d) charged impurities with $\alpha_0=1000$, and (c), (f) charged impurities with $\alpha_0=1$. Here, if the temperature is normalized by $T_{\rm F}=\varepsilon_{\rm F}/k_{\rm B}$, the result is independent of $\varepsilon_{\rm F}$ at the semi-Dirac transition point. The blue dashed-dotted lines and red dashed lines represent fitting by the corresponding asymptotic form [Eqs. (\ref{eq:asymptotic_temp_semidirac_short}) and (\ref{eq:asymptotic_temp_semidirac_long})] in the low- and high-temperature limits, respectively. 
}
\label{fig:temp_semidirac}
\end{figure}

Figure~\ref{fig:temp_semidirac} shows the temperature dependence of the dc conductivity at the semi-Dirac transition point, normalized by the zero-temperature conductivity value in each direction. For short-range impurities, we determine that the asymptotic behavior is given by 
\begin{subequations}\label{eq:asymptotic_temp_semidirac_short}
\begin{eqnarray}
{\sigma_{xx}(T)\over \sigma_{xx}(0)} &=&
\begin{cases}
1-\frac{\pi^2}{12} \left(\frac{T}{T_{\rm F}}\right)^2  & (T\ll T_{\rm F}), \\ 
\log2 \left(\frac{T}{T_{\rm F}}\right)  & (T\gg T_{\rm F}),
\end{cases} \\
{\sigma_{yy}(T)\over \sigma_{yy}(0)} &=&
\begin{cases}
1-e^{-T_{\rm F}/T}  & \!\!(T\ll T_{\rm F}), \\ 
\frac{1}{2} + \frac{1}{8\eta\left({1\over 2}\right)\Gamma\left({5\over 2}\right)} \left(\frac{T}{T_{\rm F}}\right)^{-{3\over 2}} & \!\!(T\gg T_{\rm F}).
\end{cases} 
\end{eqnarray}
\end{subequations}
For charged impurities, the asymptotic behavior is given by
\begin{subequations}\label{eq:asymptotic_temp_semidirac_long}
\begin{eqnarray}
{\sigma_{xx}(T)\over \sigma_{xx}(0)} &=&
\begin{cases}
1 + C_{xx} \left(\frac{T}{T_{\rm F}}\right)^2  & (T\ll T_{\rm F}), \\ 
D_{xx} \left(\frac{T}{T_{\rm F}}\right)^2  & (T\gg T_{\rm F}),
\end{cases} \\
{\sigma_{yy}(T)\over \sigma_{yy}(0)} &=&
\begin{cases}
1 + C_{yy} \left(\frac{T}{T_{\rm F}}\right)^2  & (T\ll T_{\rm F}), \\ 
D_{yy} \left(\frac{T}{T_{\rm F}}\right) & (T\gg T_{\rm F}),
\end{cases}
\end{eqnarray}
\end{subequations}
where $C_{ii}$ ($D_{ii}$) indicates the low- (high-) temperature coefficients. In the strong screening limit, the coefficients become $C_{xx}=0$, $D_{xx}=\frac{\pi^2}{6}$, $C_{yy}=-\frac{\pi^2}{4}$, and $D_{yy}=\log2$. As the screening strength decreases, the high-temperature coefficients $D_{ii}$ remain positive, whereas the low-temperature coefficients $C_{ii}$ decrease and we expect that the initially negative or vanishing $C_{ii}$ would eventually become positive in the weak screening limit. (See Appendix \ref{sec:app_temperature_dep_semidirac} for the detailed derivations of the coefficients $C_{ii}$ and $D_{ii}$.)

The temperature dependence in the high-temperature limit can be easily understood by replacing $\varepsilon_{\rm F}$ with $T$ in the Fermi energy dependence of dc conductivity [Eqs.~(\ref{eq:density_dependence_short_semidirac}) and (\ref{eq:density_dependence_charged_strong_screening_semidirac})]. At high temperatures, $\sigma_{yy}(T)$ for short-range impurities decreases with the temperature, showing a metallic behavior. Otherwise, the conductivities increase with the temperature, showing an insulating behavior. Note that the high-temperature asymptotic form for charged impurities is obtained by considering the effect of the energy averaging and that of the temperature-dependent screening separately. It correctly predicts the temperature power-law dependence but not the coefficients in the asymptotic form, showing a discrepancy with the numerical result, as the effect of temperature cannot be simply separated into the energy averaging and the temperature-dependent screening at high temperatures. 

\subsection{Insulator phase}

\begin{figure}[htb]
\includegraphics[width=\linewidth]{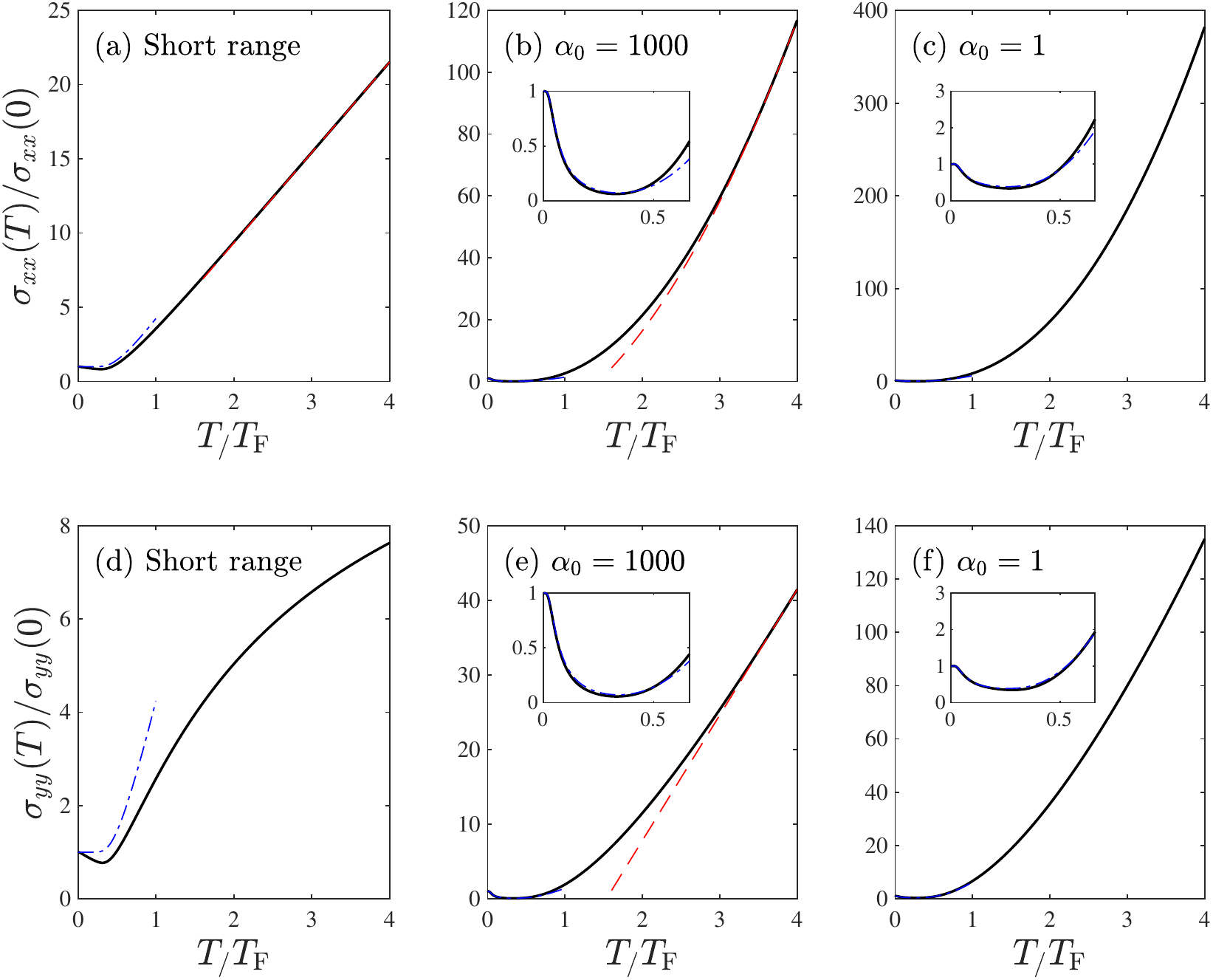}
\caption{
Calculated dc conductivities (a)-(c) $\sigma_{xx}$ and (d)-(f) $\sigma_{yy}$ in the low-density limit as a function of the temperature in the insulator phase with $\Delta=1$ for (a), (d) short-range impurities, (b), (d) charged impurities with $\alpha_0=1000$, and (c), (f) charged impurities with $\alpha_0=1$. Here, $\varepsilon_{\rm F}=1.1 \varepsilon_0$ is used for the calculation. The blue dashed-dotted lines represent the result for the gapped 2DEG system (see Appendix \ref{sec:app_temperature_dep_effmodel}), and the red dashed lines represent power-law fitting by the asymptotic form of the semi-Dirac transition point [Eqs. (\ref{eq:asymptotic_temp_semidirac_short}) and (\ref{eq:asymptotic_temp_semidirac_long})] in the high-temperature limit.
}
\label{fig:temp_semicon_110}
\end{figure}

Figure~\ref{fig:temp_semicon_110} shows the temperature dependence of the dc conductivity in the insulator phase with the fixed Fermi energy of $\varepsilon_{\rm F}=1.1 \varepsilon_0$, which corresponds to the low-density limit. At zero temperature, the insulator phase in the low-density limit can be effectively considered as a gapped 2DEG (with anisotropic effective masses). 
Similarly, at finite temperatures, the temperature-dependent conductivity of the insulator phase in the low-density limit resembles that of the gapped 2DEG system (blue dash-dotted lines in Fig.~\ref{fig:temp_semicon_110}), especially in the low-temperature limit. In the high-temperature limit, the power-law behavior of the temperature-dependent conductivity for the insulator phase becomes similar to that of the semi-Dirac transition point [Eqs. (\ref{eq:asymptotic_temp_semidirac_short}) and (\ref{eq:asymptotic_temp_semidirac_long})], because thermally excited carriers above the gap contribute to the conductivity. 
(See Appendix \ref{sec:app_temperature_dep_effmodel} for the temperature dependence of the chemical potential, Thomas--Fermi screening wave vector, and conductivity of the gapped 2DEG system.) 

In the high-density limit, the temperature dependence of dc conductivity in the insulator phase resembles that of the semi-Dirac transition point.

\subsection{Dirac semimetal phase}

\begin{figure}[htb]
\includegraphics[width=\linewidth]{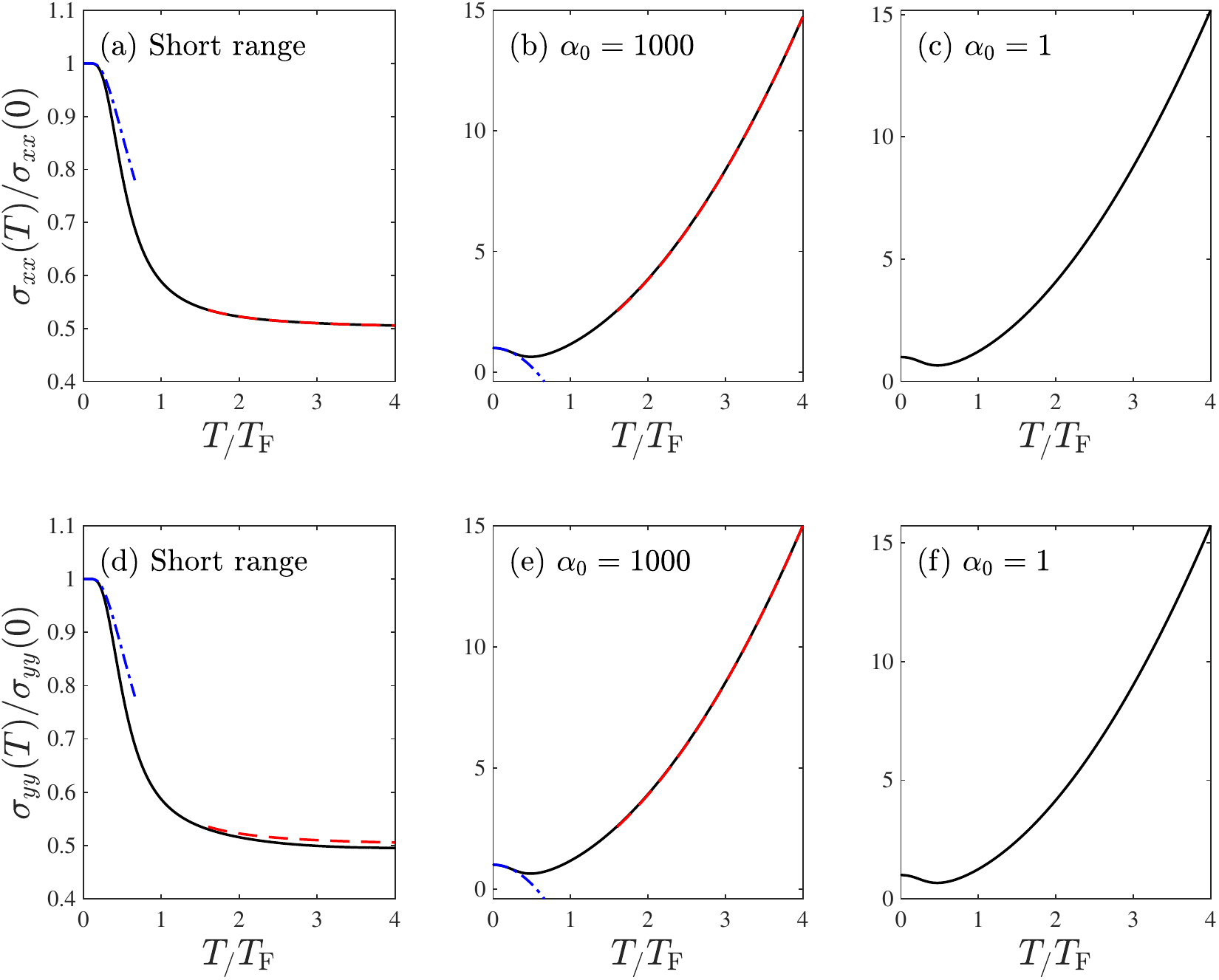}
\caption{
Calculated dc conductivities (a)-(c) $\sigma_{xx}$ and (d)-(f) $\sigma_{yy}$ in the low-density limit as a function of the temperature in the Dirac semimetal phase with $\Delta=-1$ for (a), (d) short-range impurities, (b), (d) charged impurities with $\alpha_0=1000$, and (c), (f) charged impurities with $\alpha_0=1$. Here, $\varepsilon_{\rm F}=0.01 \varepsilon_0$ is used for the calculation. The blue dashed-dotted lines and red dashed lines represent fitting by the corresponding asymptotic form [Eqs.~(\ref{eq:asymptotic_temp_dirac_short}) and (\ref{eq:asymptotic_temp_dirac_long})] in the low- and high-temperature limits, respectively. 
}
\label{fig:temp_Dirac_1}
\end{figure}

Figure~\ref{fig:temp_Dirac_1} shows the calculated temperature-dependent conductivity in the Dirac semimetal phase, with the fixed Fermi energy of $\varepsilon_{\rm F}=0.01 \varepsilon_0$, which corresponds to the low-density limit. At low densities, the Dirac semimetal phase can be effectively considered as graphene (with anisotropic velocities); thus, we can understand its temperature-dependent conductivity behavior using the result of graphene. (See Appendix \ref{sec:app_temperature_dep_effmodel} for the temperature dependence of the chemical potential, Thomas--Fermi screening wave vector, and conductivity of graphene.) For graphene with short-range impurities, the asymptotic form of the temperature-dependent conductivity becomes
\begin{eqnarray}\label{eq:asymptotic_temp_dirac_short}
{\sigma_{\rm gp}(T)\over \sigma_{\rm gp}(0)} &=&
\begin{cases}
1-e^{-T_{\rm F}/T}  & (T\ll T_{\rm F}), \\ 
\frac{1}{2}+ \frac{1}{16 \log 2} \left(\frac{T}{T_{\rm F}}\right)^{-2} & (T\gg T_{\rm F}),
\end{cases} 
\end{eqnarray}
whereas for charged impurities in the strong screening limit, the asymptotic form of the temperature-dependent conductivity becomes
\begin{eqnarray}\label{eq:asymptotic_temp_dirac_long}
{\sigma_{\rm gp}(T)\over \sigma_{\rm gp}(0)} &=&
\begin{cases}
1-\frac{\pi^2}{3} \left(\frac{T}{T_{\rm F}}\right)^{2} & (T\ll T_{\rm F}), \\ 
\frac{\pi^2}{6} \left(\frac{T}{T_{\rm F}}\right)^{2} & (T\gg T_{\rm F}).
\end{cases}
\end{eqnarray}
Similar to the result of the semi-Dirac transition point, the high-temperature asymptotic form for charged impurities correctly captures the temperature power-law dependence (but not the exact coefficient value, as discussed in Sec.~{\ref{temperature_dependence_semi-Dirac}).

\begin{figure}[htb]
\includegraphics[width=\linewidth]{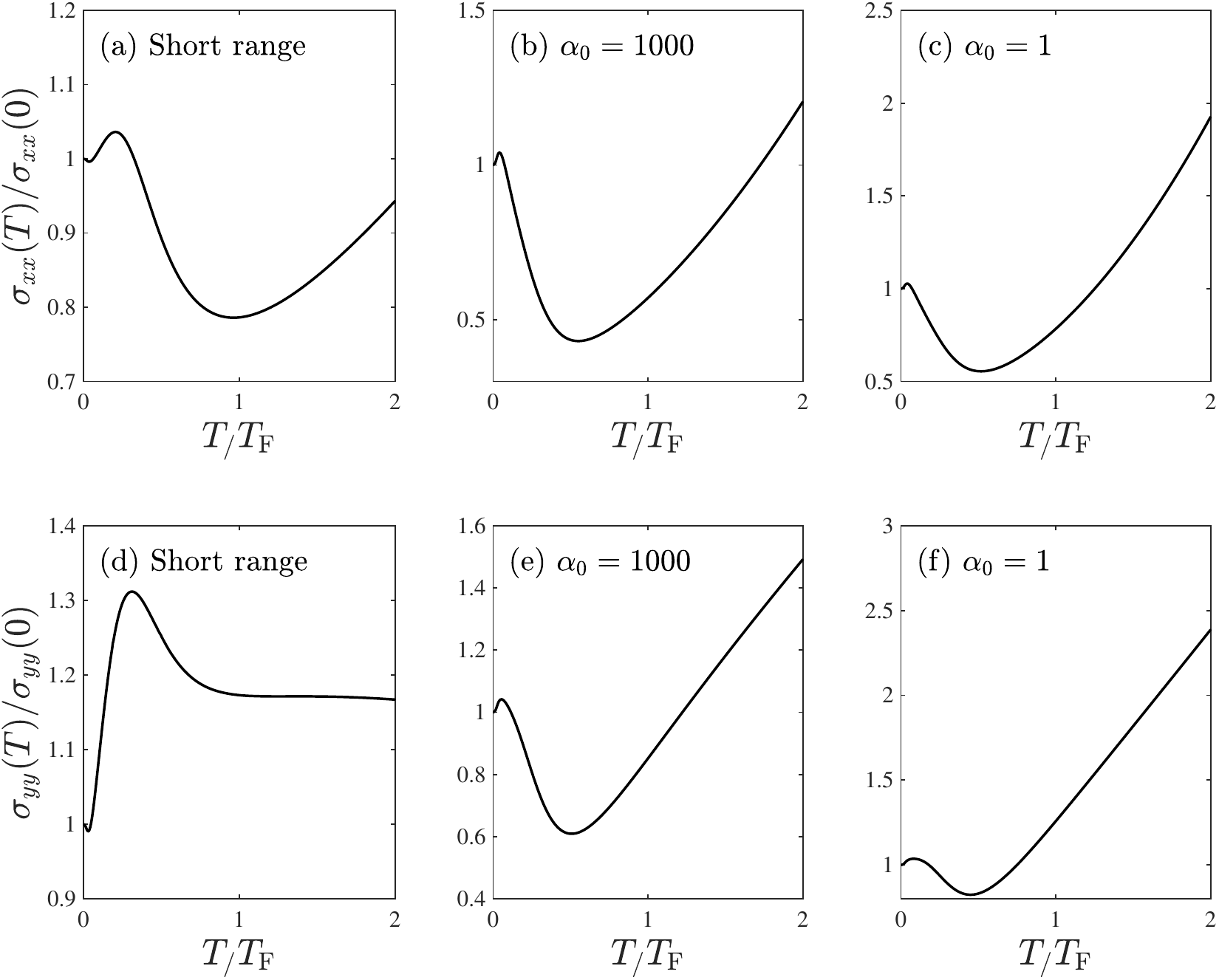}
\caption{
Calculated dc conductivities (a)--(c) $\sigma_{xx}$ and (d)--(f) $\sigma_{yy}$ immediately below the van Hove singularity point as a function of the temperature in the Dirac semimetal phase with $\Delta=1$ for (a), (d) short-range impurities, (b), (d) charged impurities with $\alpha_0=1000$, and (c), (f) charged impurities with $\alpha_0=1$. Here, $\varepsilon_{\rm F}=0.9 \varepsilon_0$ is used for the calculation.
}
\label{fig:temp_Dirac_90}
\end{figure}

Figure \ref{fig:temp_Dirac_90} shows the temperature dependence of the dc conductivity in the Dirac semimetal phase immediately below the van Hove singularity point, exhibiting a nonmonotonic behavior with temperature. As explained earlier, the temperature dependence of the dc conductivity is determined by the energy averaging with $S^{(0)}(\varepsilon)$ broadened by temperature, and by the temperature-dependent screening for charged impurities. Thus, if the Fermi energy is near the van Hove singularity, the distance between the Fermi energy and the van Hove singularity sets an important energy scale for the temperature dependence, $k_{\rm B} T_1\equiv \left||\varepsilon_{\rm F}|-|\varepsilon_{\rm g}|\right|$.
For charged impurities, the conductivity first increases, showing a peak at $T_1$, and thereafter decreases, showing a dip at $T_2^{\rm ch}\sim 0.5 T_{\rm F}$ corresponding to the minimum of $q_{\rm TF}(T)$, mainly following the temperature dependence of the screening wave vector $q_{\rm TF}(T)$ [Fig.~\ref{fig:mu_qtf_dirac}(e) in the Appendix]. For short-range impurities, the conductivity first decreases, showing a dip at $T_1$, and thereafter increases, showing a peak at $T_2^{\rm sh}\sim 0.25 T_{\rm F}$. These dips and peaks are from the temperature-dependent evolution of the chemical potential $\mu(T)$ [Fig.~\ref{fig:mu_qtf_dirac}(b) in the Appendix], shifting the central point of the energy averaging. 

In the high-density limit, the temperature dependence of dc conductivity in the Dirac semimetal phase resembles that of the semi-Dirac transition point.


\section{Discussion and conclusion}
\label{sec:discussion}
When we consider both short-range and charged impurities, assuming that each scattering mechanism is independent, the total scattering rate is obtained by adding their scattering rates in accordance with Matthiessen's rule. Note that the scattering mechanism with a higher scattering rate (or equivalently a lower conductivity) dominates the resulting conductivity. From the obtained Fermi-energy power-law dependence of dc conductivity, we can determine the dominant scattering mechanism. 
At the semi-Dirac transition point, we can observe from Eq.~(\ref{eq:density_dependence_short_semidirac}) and Figs.~\ref{fig:density_powever_evolution}(a) and (d) that, for both $\sigma_{xx}$ and $\sigma_{yy}$, the Fermi-energy power law for short-range impurities is always smaller than that of charged impurities. This indicates that, at low densities, charged impurities are dominant over short-range impurities, whereas at high densities, short-range impurities are dominant over charged impurities. 
In the insulator phase, at low densities, the system can be approximated as a 2DEG and the Fermi-energy power laws for short-range and charged impurities are almost comparable (except in the no-screening limit) as shown in Eq.~(\ref{eq:density_dependence_short_semicon}) and Figs.~\ref{fig:density_powever_evolution}(b) and (e). At high densities, the power-law dependence follows that of the semi-Dirac transition point; thus, short-range impurities dominate over charged impurities.
In the Dirac semimetal phase, at low densities, the Fermi-energy power law for short-range impurities is always smaller than that of charged impurities as shown in Eq.~(\ref{eq:density_dependence_short_Dirac}) and Figs.~\ref{fig:density_powever_evolution}(c) and (f); thus, charged impurities are dominant over short-range impurities as in the case of graphene. At high densities, short-range impurities become dominant over charged impurities, following the trend of the semi-Dirac transition point. Note that, near the van Hove singularities, charged impurities are highly screened due to the enhanced DOS, and thus, short-range impurities are dominant over charged impurities \cite{Woo2017}.


Our analysis is based on the semiclassical Boltzmann transport theory, which is known to be valid in the high-density limit. At low densities, the effect of potential fluctuations induced by spatially inhomogeneous impurities becomes important, which is not captured by our approach assuming a spatially homogeneous system. At the semi-Dirac transition point or in the Dirac semimetal phase, the potential fluctuation is expected to result in a minimum conductivity \cite{Adam18392, Li2011, Ramakrishnan2015}. In the insulator phase, if the band gap is sufficiently large, the effect of the potential fluctuation might be limited. The interplay of the impurity potential fluctuation, temperature, and band gap would be an interesting future research direction. 


Finally, we wish to mention the additional parabolic term $\gamma \frac{\hbar^2 k_y^2}{2 m^*} \sigma_x$ omitted in Eq.~(\ref{eq:BP}) along the armchair ($y$) direction beyond the lowest order \cite{Jang2018}. This term could affect the dc conductivity, especially at high densities above the crossover Fermi energy $\varepsilon_{\rm F}^{\rm cr}={2m^{\ast} v^2 \over \gamma}$, where the effective Hamiltonian in Eq.~(\ref{eq:BP}) is no longer valid.
For example, at the semi-Dirac transition point with $\varepsilon_{\rm F}\gg \varepsilon_{\rm F}^{\rm cr}$, the parabolic term becomes dominant over the linear term along the armchair direction; thus, $\sigma_{xx}$ and $\sigma_{yy}$ will follow those of (anisotropic) 2DEG.

In summary, we calculate the dc conductivity of few-layer BP as a function of the density and temperature using the anisotropic multiband Boltzmann transport theory, which is essential when the effect of anisotropic energy dispersion or interband scattering becomes important. We find that the dc conductivities in the Boltzmann limit show characteristic density and temperature dependence in each phase, which could be used as a signature of the tunable electronic structure of BP in transport measurements.


\acknowledgments
This work was supported by the National Research Foundation of Korea (NRF) grant funded by the Korea government (MSIT) (No.~2018R1A2B6007837) and Creative-Pioneering Researchers Program through Seoul National University (SNU).

\appendix

\section{Eigenstates and density of states}
\label{sec:eigenstate_dos}
In this section, we provide a detailed explanation on the model Hamiltonian of few-layer black phosphorus (BP), and its various properties including density of states (DOS). In the model Hamiltonian given by Eq.~(\ref{eq:BP}) in the main text, the exact values of $m^{*}$ and $v_0$ depend on the number of layers and the gap tuning parameter. We introduce the normalization constants $k_0 \equiv a^{-1}$ and $\varepsilon_0 \equiv \frac{\hbar^2 k_0^2}{2m^{*}}$; thus, the Hamiltonian becomes
\begin{eqnarray}
H
&=&\varepsilon_0\left(
\begin{array}{cc}
0 & \tilde{k}_{x}^2-ic\tilde{k}_{y}+\Delta\\
\tilde{k}_{x}^2+ic\tilde{k}_{y}+\Delta &0\\
\end{array}
\right),
\end{eqnarray}
where $\tilde{\bm{k}}=\bm{k}/k_0$, $c=\hbar v_0 k_0/\varepsilon_0$, and $\Delta \equiv \frac{\varepsilon_{\rm g}}{2 \varepsilon_0}$. To avoid difficulties associated with anisotropic dispersion, we consider the following coordinate transformation with
\begin{equation}\label{eq:transformation}
\begin{split}
k_x&\rightarrow \alpha k_0\left(r\cos\phi-\Delta \right)^{1\over 2},\\
k_y&\rightarrow {k_0 \over c} r\sin\phi,
\end{split}
\end{equation}
where $\alpha= \pm 1$ represents each half of the Fermi surfaces. This Fermi surface splitting is especially useful for the $\Delta<0$ case where there are two distinct Fermi surfaces (see Fig.~\ref{fig:dispersion}(f) in the main text), accounting for the ``interband'' scattering between these two surfaces. The maximum value of $\phi$ is thus given by
\begin{equation}
\phi_{\rm max}(r)=\begin{cases}
\arccos\left(\frac{\Delta}{r}\right) & (\Delta \neq 0 \text{ and } |\Delta| < r), \\
\pi & (\Delta < 0 \text{ and } |\Delta| \geq r), \\
\frac{\pi}{2} & (\text{otherwise}), \\
\end{cases}
\end{equation}
where $\phi \in \left[-\phi_{\rm max}(r),\phi_{\rm max}(r)\right]$. This coordinate transformation changes the Hamiltonian into the following form:
\begin{equation}\label{anisotropic_model}
H=\varepsilon_0 r \left(
\begin{array}{cc}
0 &e^{-i\phi}\\
e^{i\phi} &0\\
\end{array}
\right).
\end{equation}
In the transformed coordinates, the energy dispersion is given by $\varepsilon_{\pm}(r)=\pm \varepsilon_0 r$ and the corresponding eigenstates are given by
\begin{subequations}
\begin{eqnarray}
|+\rangle&=&
\frac{1}{\sqrt{2}}\left(
\begin{array}{c}
1 \\
e^{i \phi} \\
\end{array}
\right), \\
|-\rangle&=&
\frac{1}{\sqrt{2}}\left(
\begin{array}{c}
-1\\
e^{i\phi}
\end{array}
\right).
\end{eqnarray}
\end{subequations}

The Jacobian $\mathcal{J}$ corresponding to this transformation is given by
\begin{equation}\label{eq:jacobian}
\mathcal{J}=
\left|
\begin{array}{ccc}
 \frac{\partial k_x}{\partial r} & \frac{\partial k_x}{\partial \phi} \\
 \frac{\partial k_y}{\partial r}  & \frac{\partial k_y}{\partial \phi} \\
\end{array}
\right|=\frac{k_0^2r}{2c\sqrt{r \cos\phi-\Delta}} \equiv \mathcal{J}(r,\phi).
\end{equation}
Note that, for the $+$ band, the band velocity $v_{\bm k}^{(i)}={1\over \hbar} {\partial\varepsilon_{+,\bm{k}} \over \partial  k_i}$ can be expressed as
\begin{subequations}\label{eq:band_velocity}
\begin{eqnarray}
v_{\bm k}^{(x)}&=&2\alpha v_0\cos\phi\sqrt{r\cos\phi-\Delta}, \\
v_{\bm k}^{(y)}&=& v_0 c \sin\phi,
\end{eqnarray}
\end{subequations}
where $v_0={\varepsilon_0\over \hbar k_0}$.

The DOS at the semi-Dirac transition point ($\Delta=0$) at the energy $\varepsilon>0$ can be obtained analytically as
\begin{eqnarray}\label{eq:dos_bp}
D(\varepsilon)&=&g\int {d^2k\over (2\pi)^2} \delta(\varepsilon-\varepsilon_{+,\bm{k}}) \nonumber \\
&=&2 g \int_0^{\infty}dr \int_{- {\pi \over 2}}^{\pi \over 2}d\phi {\mathcal{J}(r,\phi)\over (2\pi)^2}\delta(\varepsilon-\varepsilon_0 r)  \nonumber \\
&=&{2 g k_0^2 \sqrt{2} K(1/2)\over \pi^2 c \varepsilon_0}\left(\varepsilon\over \varepsilon_0\right)^{1\over 2},
\end{eqnarray}
where $g$ is the spin degeneracy, and the factor 2 originates from the duplicate parts of the Fermi surfaces parameterized by $\alpha=\pm 1$.
Here, $K(k)=\sum_{n=0}^{\infty} \left[(2n-1)!!/(2n)!!\right]^2k^{2n}$ is the complete elliptic integral of the first kind with $K(1/2)\approx 1.854$ \cite{Arfken2012}.
Note that the Thomas--Fermi wave vector is determined by the DOS at the Fermi energy $\varepsilon_{\rm F}$ given by
\begin{equation}\label{eq:q_TF_mWSM}
q_{\rm TF}={2\pi e^2 \over \kappa} D(\varepsilon_{\rm F})={4 g \alpha_0 k_0 \sqrt{2} K(1/2)\over \pi c} \left(\frac{\varepsilon_{\rm F}}{\varepsilon_0} \right)^{1\over 2},
\end{equation}
where $\alpha_0={e^2\over \kappa \hbar v_0}$ is the effective fine structure constant.

The carrier density is thus given by
\begin{equation}
n=\int_0^{\varepsilon_{\rm F}}d\varepsilon D(\varepsilon)=n_0{4g  \sqrt{2} K(1/2)\over 3 \pi^2 c} \left(\frac{\varepsilon_{\rm F}}{\varepsilon_0} \right)^{3\over 2},
\end{equation}
where $n_0=k_0^2$.
Note that $\varepsilon_{\rm F}\sim n^{2\over  3}$ and $D(\varepsilon_{\rm F})\sim n^{1\over 3}$.

Figure~\ref{fig:DOS_density} in the main text shows the calculated DOS and the carrier density for each phase.


\section{Density dependence of dc conductivity in black phosphorus}
\label{sec:conductivity_2D_anisotropic}

In this section, we derive the dc conductivity at zero temperature for 2D multiband systems with anisotropic energy dispersion. To consider the anisotropy of the energy dispersion, we express the multiband anisotropic Boltzmann equation in Eq.~(\ref{eq:relaxation_time_anisotropic}) using the transformed coordinates in Eq.~(\ref{eq:jacobian}) as follows:
\begin{widetext}
\begin{eqnarray}\label{eq:relaxation_time_anisotropic_theta}
1&=&\sum_{\alpha'}\int_0^{\infty}dr'\int_{-\phi_{\rm max}(r')}^{\phi_{\rm max}(r')}d\phi'{\mathcal{J}(r',\phi')\over (2\pi)^2} W_{\bm{k}\bm{k}'}^{\alpha \alpha'} \left(\tau_{\bm{k} \alpha}^{(i)}-{v_{\bm{k}'\alpha'}^{(i)}\over v_{\bm{k}\alpha}^{(i)} }\tau_{\bm{k}'\alpha'}^{(i)}\right) \nonumber \\
&=&\sum_{\alpha'} \int_0^{\infty}dr'\int_r' \frac{d\phi'}{(2\pi)^2}
\frac{ k_0^2r'}{2c\sqrt{r' \cos\phi'-\Delta}} 
\left[{2\pi\over\hbar} n_{\rm imp} |V_{\bm{k}\bm{k}'}^{\alpha \alpha'}|^2 F_{\bm{k}\bm{k}'}^{\alpha \alpha'} \delta(\varepsilon_0 r-\varepsilon_0 r')\right]\left(\tau_{\bm{k}\alpha}^{(i)}-d_{\bm{k}\bm{k}' }^{\alpha \alpha'(i)}\tau_{\bm{k}' \alpha'}^{(i)}\right) \nonumber \\
&=&{2\pi\over\hbar} n_{\rm imp} {\frac{k_0^2}{2\pi c \varepsilon_0}} \sum_{\alpha'} \int_r'{d\phi' \over 2\pi}\frac{r}{2 \sqrt{r \cos\phi'-\Delta}} |V_{\bm{k}\bm{k}'}^{\alpha \alpha'}|^2 F_{\bm{k}\bm{k}'}^{\alpha \alpha'} \left(\tau_{\bm{k}\alpha}^{(i)}-d_{\bm{k}\bm{k}' }^{\alpha \alpha'(i)}\tau_{\bm{k}' \alpha'}^{(i)}\right),
\end{eqnarray}
\end{widetext}
where $\alpha=\pm 1$ represents each half of the Fermi surfaces, $d_{\bm{k}\bm{k}' }^{\alpha \alpha'(i)}=v_{\bm{k}'\alpha'}^{(i)}/v_{\bm{k}\alpha}^{(i)} $, and $F_{\bm{k}\bm{k}'}^{\alpha \alpha'} =  \frac{1}{2} \left[1+\cos (\phi - \phi')\right]$ is the square of the wave function overlap between $\bm{k}$ and $\bm{k}'$ states in the same conduction (or valence) band. 
Let us define $\rho_0={k_0^2 \over 2\pi  c \varepsilon_0}$, $V_0={\varepsilon_0\over k_0^2}$, and ${1\over \tau_0(r)}={2\pi\over \hbar} n_{\rm imp} V_0^2 \rho_0$, then we have
\begin{eqnarray}\label{eq:relaxation_time_anisotropic_mu}
&&1=\sum_{\alpha'} \int_r'{d\phi' \over 2\pi}\frac{r}{2 \sqrt{r \cos\phi'-\Delta}} \nonumber \\
&& \times |\tilde{V}_{\bm{k}\bm{k}'}^{\alpha \alpha'}|^2 F_{\bm{k}\bm{k}'}^{\alpha \alpha'} \left(\tilde{\tau}_{\bm{k}\alpha}^{(i)}-d_{\bm{k}\bm{k}' }^{\alpha \alpha'(i)}\tilde{\tau}_{\bm{k}' \alpha'}^{(i)}\right),
\end{eqnarray}
where $\tilde{V}_{\bm{k}\bm{k}'}^{\alpha \alpha'(i)}=V_{\bm{k}\bm{k}'}^{\alpha \alpha'(i)}/V_0$ and $\tilde{\tau}_{\bm{k} \alpha}^{(i)}=\tau_{\bm{k} \alpha}^{(i)}/\tau_0$. Here, $\int_r' d\phi'$ represents an integration over $-\phi_{\rm max}(r)<\phi'<\phi_{\rm max}(r)$.
Thus, Eq.~(\ref{eq:relaxation_time_anisotropic_mu}) becomes
\begin{equation}\label{eq:Boltzmann_BP}
1=\sum_{\alpha'} \left [\tilde{w}^{(i)}_{\alpha \alpha'}(\phi)\tilde{\tau}_{\alpha}^{(i)}(\phi)-\int_r'\frac{d\phi'}{2\pi} \tilde{w}^{(i)}_{\alpha \alpha'}(\phi,\phi')\tilde{\tau}_{\alpha'}^{(i)}(\phi') \right], 
\end{equation}
where
\begin{subequations}
\label{eq:scattering_rate_matrix}
\begin{eqnarray}
\tilde{w}^{(i)}_{\alpha \alpha'}(\phi)\!&=&\! \int_r'\! {d\phi' \over 2\pi}\frac{r}{2\sqrt{r \cos\phi'-\Delta}}   |V_{\bm{k}\bm{k}'}^{\alpha \alpha'}|^2 F_{\bm{k}\bm{k}'}^{\alpha \alpha'}, \nonumber \\ \\
\tilde{w}^{(i)}_{\alpha \alpha'}(\phi,\phi')\!&=&\! \frac{r}{2\sqrt{r \cos\phi'-\Delta}}|V_{\bm{k}\bm{k}'}^{\alpha \alpha'}|^2 F_{\bm{k}\bm{k}'}^{\alpha \alpha'} d_{\bm{k}\bm{k}' }^{\alpha \alpha'(i)}. \nonumber \\
\end{eqnarray}
\end{subequations}
Since Eq.~(\ref{eq:Boltzmann_BP}) holds for both $\alpha=\pm1$, we can rewrite it as
\begin{subequations}
\label{eq:Boltzmann_BP_expand}
\begin{eqnarray}
1&=&\tilde{w}^{(i)}_{11}(\phi)\tilde{\tau}_{1}^{(i)}(\phi)-\int_r' \frac{d\phi'}{2\pi} \tilde{w}^{(i)}_{11}(\phi,\phi')\tilde{\tau}_{1}^{(i)}(\phi') \\ \nonumber
&+& \tilde{w}^{(i)}_{1-1}(\phi)\tilde{\tau}_{1}^{(i)}(\phi)-\int_r' \frac{d\phi'}{2\pi} \tilde{w}^{(i)}_{1-1}(\phi,\phi')\tilde{\tau}_{-1}^{(i)}(\phi'), \\
1&=&\tilde{w}^{(i)}_{-11}(\phi)\tilde{\tau}_{-1}^{(i)}(\phi)-\int_r' \frac{d\phi'}{2\pi} \tilde{w}^{(i)}_{11}(\phi,\phi')\tilde{\tau}_{1}^{(i)}(\phi') \\ \nonumber
&+& \tilde{w}^{(i)}_{-1-1}(\phi)\tilde{\tau}_{-1}^{(i)}(\phi)-\int_r' \frac{d\phi'}{2\pi} \tilde{w}^{(i)}_{-1-1}(\phi,\phi')\tilde{\tau}_{-1}^{(i)}(\phi').
\end{eqnarray}
\end{subequations}

Now, let us discretize $\phi$ to $\phi_n$ ($n=1,2,\cdots,N$) with an interval $\Delta\phi=2\phi_{\rm max}(r)/N$. Thus, for $\tilde{\tau}_{n \alpha}^{(i)}=\tilde{\tau}^{(i)}_{\alpha}(\phi_n)$, we have
\begin{subequations}
\label{eq:relaxation_time_matrix_form}
\begin{eqnarray}
1&=&P^{(i)}\substack{n \\ 1 1} \tilde{\tau}_{n 1}^{(i)} - \sum_{n'} P^{(i)}\substack{nn' \\ 11}\tilde{\tau}_{n' 1}^{(i)} \nonumber \\
&+&P^{(i)}\substack{n \\ 1 -1 } \tilde{\tau}_{n 1}^{(i)} - \sum_{n'} P^{(i)}\substack{nn' \\ 1-1 }\tilde{\tau}_{n' -1}^{(i)}, \\
1&=&P^{(i)}\substack{n \\ -1 1} \tilde{\tau}_{n -1}^{(i)} - \sum_{n'} P^{(i)}\substack{nn' \\ -11}\tilde{\tau}_{n' 1}^{(i)} \nonumber \\ 
&+&P^{(i)}\substack{n \\ -1 -1 } \tilde{\tau}_{n -1}^{(i)} - \sum_{n'} P^{(i)}\substack{nn' \\ -1-1 }\tilde{\tau}_{n' -1}^{(i)}, 
\end{eqnarray}
\end{subequations}
where $P^{(i)}\substack{n \\ \alpha \alpha' } =\tilde{w}^{(i)}_{\alpha \alpha'}(\phi_n)$ is an $N$-vector and $P^{(i)}\substack{nn' \\ \alpha \alpha'}=\tilde{w}^{(i)}_{\alpha \alpha'}(\phi_n,\phi_{n'})\Delta\phi$ is an $N\times N$ matrix, which correlates the different $\phi$-dependent relaxation times for a given $(\alpha$, $\alpha')$ combination. Note that Eq.~(\ref{eq:relaxation_time_matrix_form}) shares the basic structure with the multiband scattering formula \cite{Siggia1970,Woo2017} (which accounts for the scattering between each half of the Fermi surface) and the anisotropic scattering formula \cite{Park2017} (which accounts for the scattering between different $\phi$ and $\phi'$ points). Furthermore,  Eq.~(\ref{eq:relaxation_time_matrix_form}) is a $2N\times2N$ matrix equation with two independent basis indices ($\alpha$, $\phi_n$), i.e., index  $\alpha$ for each half of the Fermi surfaces and the $\phi$-discretization index $n$. 

Thus, the dc conductivity at zero temperature is given by
\begin{eqnarray}
\label{eq:conductivity_zero_temperature}
&&\sigma_{ij}=g e^2\sum_{\alpha}\int {d^2 k\over (2\pi)^2} \delta(\varepsilon_{{\bm k}}-\varepsilon_{\rm F}) v_{\bm{k}\alpha}^{(i)} v_{\bm{k}\alpha}^{(j)}\tau_{\bm{k}\alpha}^{(j)} \nonumber \\
&&=g e^2 \sum_{\alpha} \int_0^{\infty}dr \int_r'd\phi \frac{ k_0^2r \delta(\varepsilon_0 r-\varepsilon_{\rm F})v_{\bm{k}\alpha}^{(i)} v_{\bm{k}\alpha}^{(j)}\tau_{\bm{k}\alpha}^{(j)}}{2(2\pi)^2c\sqrt{r \cos\phi-\Delta}} \nonumber \\
&&=\sigma_0\sum_{\alpha} \int_0^{\infty}dr \int_r' {d\phi \over 2\pi} \frac{ r \delta(r-r_{\rm F}) \tilde{v}_{\bm{k}\alpha}^{(i)} \tilde{v}_{\bm{k}\alpha}^{(j)} \tilde{\tau}_{\bm{k}\alpha}^{(j)} }{2\sqrt{r \cos\phi-\Delta}},
\end{eqnarray}
where $\sigma_0=g e^2 \rho_0 v_0^2 \tau_0$, $r_{\rm F}=\varepsilon_{\rm F}/\varepsilon_0$ and $\tilde{v}_{\bm{k}\alpha}^{(i)}=v_{\bm{k}\alpha}^{(i)}/v_0$.
Thus, from Eq.~(\ref{eq:band_velocity}), we have
\begin{subequations}\label{eq:sigma_xx_sigma_zz}
\begin{eqnarray}
{\sigma_{xx} \over \sigma_0} &=& 2  \sum_{\alpha}\int_{r_{\rm F}}'\frac{d\phi}{2\pi} r_{\rm F} \cos^2\phi \sqrt{r_{\rm F}  \cos\phi-\Delta} \tilde{\tau}_{\alpha}^{(x)}(\phi), \nonumber \\ \\
{\sigma_{yy} \over \sigma_0} &=& c^2 \sum_{\alpha} \int_{r_{\rm F}}'\frac{d\phi}{2\pi} \frac{r_{\rm F} \sin^2\phi}{2\sqrt{r_{\rm F}  \cos\phi-\Delta}}  \tilde{\tau}_{\alpha}^{(y)}(\phi).
\end{eqnarray}
\end{subequations}
Note that $\tau_0$, $v_0$, $\rho_0$, and $\sigma_0$ are the density-independent normalization constants in units of time, velocity, DOS, and conductivity, respectively.

\section{Low-density approximate models for the insulator phase and Dirac semimetal phase}
\label{sec:eff_model_semicon_Dirac}

In this section, we derive the dc conductivity of low-density approximate models for the insulator phase and Dirac semimetal phase. Note that the only anisotropy considered in these models is the anisotropy in the effective mass or velocity with the \emph{same} power-law dependence in momentum.

\subsection{Insulator phase at low densities}
\label{subsec:eff_semicon}

For the insulator phase, as well as the Dirac semimetal phase discussed later, the DOS and carrier density do not follow the simple power-law behavior. Therefore, we utilize approximate models to understand the asymptotic behavior of dc conductivity at low densities. When $|\varepsilon_{\rm F}|>|\varepsilon_{\rm g}|$ but the carrier density is sufficiently small, the system can be approximated as a two-dimensional electron gas (2DEG). From the series expansion at the minimum point of the conduction band, we have
\begin{eqnarray}
\varepsilon(\bm{k})&=&\varepsilon_0 \left( k_x \over k_0 \right)^2 + \frac{\varepsilon_0 c^2}{2 \Delta} \left( k_y \over k_0 \right)^2 \nonumber \\ 
&\equiv & \frac{\hbar^2 k_x^2 }{2m_x}+ \frac{\hbar^2 k_y^2 }{2m_y},
\end{eqnarray}
where $m_x=\frac{\hbar^2 k_0^2}{2\varepsilon_0}$ and $m_y=\frac{\Delta \hbar^2 k_0^2}{c^2 \varepsilon_0}$. 

For comparison, we first consider a 2DEG with an isotropic energy dispersion given by
\begin{eqnarray}
\varepsilon(\bm{k}) = \frac{\hbar^2 k^2}{2 m}.
\end{eqnarray}
As the system is isotropic, we can readily calculate the conductivity of each case using the Einstein relation 
\begin{equation}
\label{eq:einstein_relation}
\sigma_{\rm{iso}}=e^2 D(\varepsilon_{\rm{F}}) \mathcal{D}, 
\end{equation}
where $\mathcal{D}=\frac{v_{\rm{F}}^2 \tau_{\rm{F}}}{2}$ is the diffusion constant and $D(\varepsilon) = \frac{g m}{2 \pi \hbar^2}$ is the DOS for the isotropic 2DEG.
The relaxation time at the Fermi energy $\tau_{\rm F}$ is given by
\begin{eqnarray}
\frac{1}{\tau_{\rm{F}}}&=& \frac{2\pi n_{\rm{imp}}}{\hbar}  \int{\frac{d^2\bm{k}'}{(2\pi)^2}}|V_{\bm{k}\bm{k}'}|^2 \delta(\varepsilon -\varepsilon_{\rm{F}})(1-\cos \phi') \nonumber \\
&=& \frac{2\pi n_{\rm{imp}}}{\hbar} \frac{m}{2\pi \hbar^2} \int_0^{2\pi}{\frac{d\phi'}{(2\pi)}}|V_{\phi'}|^2 (1-\cos \phi') \nonumber \\
&\equiv& \frac{2\pi n_{\rm{imp}}}{\hbar} \frac{m}{2\pi \hbar^2} \bar{V}^{2}_{\rm i2DEG},
\end{eqnarray} 
where $V_{\phi'}$ is the angle-dependent potential on the Fermi surface and $\bar{V}^{2}_{\rm i2DEG} \equiv \int_0^{2\pi}{\frac{d\phi'}{2\pi}}|V_{\phi'}|^2 (1-\cos \phi')$ is the angle-averaged square of the impurity potential. 

Therefore, the dc conductivity of the isotropic 2DEG is given by
\begin{eqnarray}
\sigma_{\rm{iso}}&=&e^2 \left(\frac{g m}{2 \pi \hbar^2}\right) \left( \frac{v_{\rm{F}}^2}{2} \right) \left( \frac{\hbar}{2\pi n_{\rm{imp}}} \frac{2\pi \hbar^2}{m \bar{V}^{2}_{\rm i2DEG}}\right) \nonumber \\
&=&\frac{ge^2 \hbar}{2\pi n_{\rm{imp}}\bar{V}^{2}_{\rm i2DEG}}\left( \frac{\hbar^2 k_{\rm{F}}^2}{2 m^2} \right) \nonumber \\
&=&\frac{ge^2 \hbar \varepsilon_{\rm{F}}}{2\pi n_{\rm{imp}}m\bar{V}^{2}_{\rm i2DEG}},
\end{eqnarray}
where $v_{\rm F}=\frac{\hbar k_{\rm F}}{m}$ and $\varepsilon_{\rm F}=\frac{\hbar^2 k_{\rm{F}}^2}{2 m^2}$.

Subsequently, let us consider the Fermi energy dependence of the dc conductivity using the Einstein relation in Eq.~(\ref{eq:einstein_relation}). For short-range impurities, $\bar{V}^{2}_{\rm i2DEG}$ is a constant independent of $\varepsilon_{\rm F}$; thus, we have
\begin{eqnarray}
\label{eq:app_density_dependence_short_semicon}
\sigma&\sim& \varepsilon_{\rm F}^{}.
\end{eqnarray}
Here, we used $v_{\rm F}^2\sim k_{\rm F}^2\sim \varepsilon_{\rm F}$.
For charged impurities in the strong screening limit, $\bar{V}^{2}_{\rm i2DEG}\sim q_{\rm TF}^{-2}\sim D^{-2}(\varepsilon_{\rm F})$ is also a constant; thus,
\begin{eqnarray}
\label{eq:app_density_dependence_charged_strong_screening_semicon}
\sigma&\sim& \varepsilon_{\rm F}^{}.
\end{eqnarray}

For the anisotropic 2DEG with different effective masses in each direction, we introduce the following coordinate transformation [$(k_x, k_y)\rightarrow(k, \phi)$]:
\begin{equation}
\begin{split}
k_x&\rightarrow \sqrt{\frac{m_x}{m}} k \cos \phi,\\
k_y&\rightarrow \sqrt{\frac{m_x}{m}} k \sin \phi,
\end{split}
\end{equation}
which gives the Jacobian $dk_x dk_y = \frac{\sqrt{m_x m_y}}{m} k dk d\phi$. 
The band velocity $v_{\bm{k}}^{(i)}=\frac{1}{\hbar}\frac{\partial\varepsilon_{\bm{k}}}{\partial k_i}$ can be expressed as
\begin{equation}
\begin{split}
v_{\bm{k}}^{(x)}=\frac{\hbar k}{\sqrt{m m_x}}\cos \phi, \\
v_{\bm{k}}^{(y)}=\frac{\hbar k}{\sqrt{m m_y}}\sin \phi.
\end{split}
\end{equation}

Subsequently, the energy dispersion becomes isotropic in the transformed coordinates; thus, the DOS is given by
\begin{eqnarray}
D(\varepsilon) = \frac{g \sqrt{m_x m_y}}{2 \pi \hbar^2}.
\end{eqnarray}

The relaxation time of the anisotropic 2DEG for $\bm k$ at the Fermi energy can be obtained by solving the coupled integral equation [Eq.~(\ref{eq:relaxation_time_anisotropic}) in the main text]. For short-range impurities or charged impurities in the strong screening limit, the scattering potential $V_{\bm{k}\bm{k}'}=V_0$ is independent of the angle, thus it can be shown that $\tau_{\bm k}^{(i)}=\tau^{(i)}_{\varepsilon_{\bm k}} \equiv \tau^{(i)}$. Then the coupled equation can be simplified as
\begin{eqnarray}\label{eq:tau_2DEG}
\frac{1}{{\tau}^{(i)}}&=& \frac{2\pi n_{\rm{imp}}}{\hbar}\!\!\int{\frac{d^2\bm{k}'}{(2\pi)^2}}|V_{\bm{k}\bm{k}'}|^2 \delta(\varepsilon_{\bm k}-\varepsilon_{\bm k'})\left(1-\frac{v_{\bm{k}'}^{(i)}}{v_{\bm{k}}^{(i)}}\right) \nonumber \\
&=& \frac{2\pi n_{\rm{imp}}}{\hbar}\frac{\sqrt{m_x m_y}}{2 \pi \hbar^2} \bar{V}^{2}_{\rm a2DEG},
\end{eqnarray}
where $\bar{V}^{2}_{\rm a2DEG} \equiv \int_0^{2\pi}{\frac{d\phi'}{(2\pi)}}|V_0|^2 \left(1-\frac{v_{\bm{k}'}^{(i)}}{v_{\bm{k}}^{(i)}}\right)=|V_0|^2$ is the angle-averaged square of the impurity potential for the anisotropic 2DEG. Note that $\tau^{(i)}$ is independent of the direction $i$. 

Therefore, the conductivity of the anisotropic 2DEG is given by
\begin{eqnarray}
\sigma_{ij}&=&g e^2 \int{\frac{d^2\bm{k}}{(2\pi)^2}} \delta(\varepsilon_{\rm{F}}-\varepsilon(\bm{k}))v^{(i)} v^{(j)} \tau ^{(j)} \nonumber \\
&=& \frac{g e^2 \sqrt{m_x m_y}}{2 \pi \hbar^2}\tau_{\rm{F}} \int_0^{2\pi}{\frac{d\phi}{2\pi}} v_{\rm{F}}^{(i)} v_{\rm{F}}^{(j)},
\end{eqnarray}
where $\tau_{\rm{F}}$ is the relaxation time at the Fermi energy.
When the electric field and the current density are along the $x$-direction, the conductivity $\sigma_{xx}$ becomes 
\begin{eqnarray}
\label{eq:sigma_xx_2DEG}
&\sigma_{xx}& = \frac{g e^2 \sqrt{m_x m_y}}{2 \pi \hbar^2}\tau_{\rm{F}} \int_0^{2\pi}{\frac{d\phi}{2\pi}}\left[v_{\rm{F}}^{(x)}\right]^2 \nonumber \\
&=&\frac{g e^2 \sqrt{m_x m_y}}{2 \pi \hbar^2}  \frac{\hbar} {2\pi n_{\rm{imp}}}\frac{2\pi \hbar^2}{\sqrt{m_x m_y} \bar{V}^2_{\rm a2DEG}} \nonumber \\
&&\times\frac{\hbar^2}{m m_x}\int_0^{2\pi} {\frac{d\phi}{2\pi}} k_{\rm{F}}^2 \cos^2 \phi \nonumber \\ 
&=& \frac{g e^2 \hbar \varepsilon_{\rm{F}}}{2\pi n_{\rm{imp}}  \bar{V}^2_{\rm a2DEG} m} \frac{m}{m_x} .
\end{eqnarray}
Similarly, when the electric field and the current density are along the $y$-direction, the conductivity $\sigma_{yy}$ becomes 
\begin{eqnarray}
\label{eq:sigma_yy_2DEG}
\sigma_{yy} &=& \frac{g e^2 \sqrt{m_x m_y}}{2 \pi \hbar^2}\tau_{\rm{F}} \int_0^{2\pi}{\frac{d\phi}{2\pi}}\left[v_{\rm{F}}^{(y)}\right]^2 \nonumber \\
&=& \frac{g e^2 \hbar \varepsilon_{\rm{F}}}{2\pi n_{\rm{imp}}  \bar{V}^2_{\rm a2DEG} m} \frac{m}{m_y}.
\end{eqnarray}
Therefore, the dc conductivities for the anisotropic case are modified as
\begin{subequations}
\begin{eqnarray}
\sigma_{xx} = \sigma_{\rm{iso}} \frac{m}{m_x}, \\ 
\sigma_{yy} = \sigma_{\rm{iso}} \frac{m}{m_y}. 
\end{eqnarray}
\end{subequations}
Thus, for short-range impurities or charged-impurities in the strong screening limit, the Fermi energy  dependence of the dc conductivities for the anisotropic 2DEG follows that of the isotropic 2DEG given by Eqs.~(\ref{eq:app_density_dependence_short_semicon}) and (\ref{eq:app_density_dependence_charged_strong_screening_semicon}).

Note that, as the Fermi energy or the carrier density increases, the insulator phase can no longer be approximated by a 2DEG model, and the energy dispersion follows that of the semi-Dirac transition point. Therefore, the power-law dependence eventually follows that of the semi-Dirac transition point.

\subsection{Dirac semimetal phase at low densities}
\label{subsec:eff_dirac}

For the Dirac semimetal phase ($\Delta<0$), the series expansion at one of the band touching points gives
\begin{eqnarray}
H(\bm{k})&=& \frac{\varepsilon_0}{k_0} \left(2 \sqrt{-\Delta} k_x \sigma_x + c k_y \sigma_y \right) \\ \nonumber
&\equiv&\hbar \left(v_x k_x \sigma_x + v_y k_y \sigma_y \right),
\end{eqnarray}
where $v_x = \frac{2 \sqrt{-\Delta} \varepsilon_0}{\hbar k_0}$ and $v_y = \frac{c \varepsilon_0}{\hbar k_0}$.

For comparison, we first consider an isotropic 2D Dirac semimetal with the Hamiltonian given by
\begin{eqnarray}
H(\bm{k})=\hbar v \left(k_x \sigma_x + k_y \sigma_y \right).
\end{eqnarray}
The DOS is thus given by
\begin{eqnarray}
D(\varepsilon) = \frac{g k}{2 \pi \hbar v} = \frac{g \varepsilon}{2 \pi \hbar^2 v^2}.
\end{eqnarray}
The relaxation time at the Fermi energy $\tau_{\rm F}$ is given by
\begin{eqnarray}
\frac{1}{\tau_{\rm{F}}}&=& \frac{2\pi n_{\rm{imp}}}{\hbar}  \int{\frac{d^2\bm{k}'}{(2\pi)^2}}|V_{\bm{k}\bm{k}'}|^2 F_{\bm{k}\bm{k}'} \delta(\varepsilon -\varepsilon_{\rm{F}})(1-\cos \phi')  \nonumber \\
&=& \frac{2\pi n_{\rm{imp}}}{\hbar} \frac{k_{\rm{F}}}{2\pi \hbar v}\int_0^{2\pi}{\frac{d\phi'}{2\pi}}|V_{\phi'}|^2 F(\phi') (1-\cos \phi') \nonumber \\
&=& \frac{2\pi n_{\rm{imp}}}{\hbar} \frac{k_{\rm{F}}}{2\pi \hbar v}\bar{V}^{2}_{\rm igp},
\end{eqnarray}
where  $F(\phi')= \frac{1}{2}(1+\cos \phi')$ is the square of the wave function overlap and $\bar{V}^{2}_{\rm igp} \equiv \int_0^{2\pi}{\frac{d\phi'}{2\pi}}|V_{\phi'}|^2 F(\phi') (1-\cos \phi')$ is the angle-averaged square of the impurity potential. 
Therefore, the dc conductivity of the isotropic Dirac semimetal is given by
\begin{eqnarray}
\sigma_{\rm{iso}}&=&e^2 \left(\frac{g k_{\rm{F}}}{2 \pi \hbar v}\right) \frac{v^2}{2} \left( \frac{\hbar}{2\pi n_{\rm{imp}}} \frac{2\pi \hbar v}{k_{\rm{F}}\bar{V}^{2}_{\rm igp}}\right) \nonumber \\
&=& \frac{ge^2 \hbar v^2}{4\pi n_{\rm{imp}}\bar{V}^{2}_{\rm igp}}.
\end{eqnarray}

Subsequently, let us consider the Fermi energy dependence of the dc conductivity using the Einstein relation in Eq.~(\ref{eq:einstein_relation}).
For short-range impurities, $\bar{V}^{2}_{\rm igp}$ is a constant independent of $\varepsilon_{\rm F}$; thus, we have
\begin{eqnarray}
\label{eq:app_density_dependence_short_Dirac}
\sigma&\sim& \varepsilon_{\rm F}^{0},
\end{eqnarray}
whereas for charged impurities in the strong screening limit, $\bar{V}^{2}_{\rm igp}\sim q_{\rm TF}^{-2}\sim D^{-2}(\varepsilon_{\rm F})\sim\varepsilon_{\rm F}^{-2}$; thus,
\begin{eqnarray}
\label{eq:app_density_dependence_charged_strong_screening_Dirac}
\sigma&\sim& \varepsilon_{\rm F}^{2}.
\end{eqnarray}
Note that, even in the weak screening limit, $\bar{V}^{2}_{\rm igp}\sim k_{\rm F}^{-2}\sim\varepsilon_{\rm F}^{-2}$, and in general, $\sigma\sim \varepsilon_{\rm F}^{2}$ for charged impurities.

For the anisotropic Dirac semimetals with different velocities in each direction, we introduce the following coordinate transformation [$(k_x, k_y)\rightarrow(k, \phi)$]:
\begin{equation}
\begin{split}
k_x&\rightarrow \frac{v}{v_x} k \cos \phi,\\
k_y&\rightarrow \frac{v}{v_y} k \sin \phi,
\end{split}
\end{equation}
which gives the Jacobian $dk_x dk_y = \frac{v^2}{v_x v_y} k dk d\phi$.
The band velocity $v_{\bm{k}}^{(i)}=\frac{1}{\hbar}\frac{\partial\varepsilon_{\bm{k}}}{\partial k_i}$ can be expressed as
\begin{equation}
\begin{split}
v_{\bm{k}}^{(x)}=v_{x}\cos \phi, \\
v_{\bm{k}}^{(y)}=v_{y}\sin \phi.
\end{split}
\end{equation}

Subsequently, the energy dispersion becomes isotropic in the transformed coordinates; thus, the DOS is given by
\begin{eqnarray}
D(\varepsilon) = \frac{g v k}{2 \pi \hbar v_x v_y} = \frac{g \varepsilon }{2 \pi \hbar^2 v_x v_y}.
\end{eqnarray}
Similarly, using the same assumptions which were used in Eq.~(\ref{eq:tau_2DEG}), for short-range impurities or charged impurities in the strong screening limit, we can calculate the relaxation time of the anisotropic Dirac semimetals given by
\begin{eqnarray}
\frac{1}{\tau^{(i)}}&=& \frac{2\pi n_{\rm{imp}}}{\hbar}\!\!  \int{\frac{d^2\bm{k}'}{(2\pi)^2}}|V_{\bm{k}\bm{k}'}|^2 F_{\bm{k}\bm{k}'} \delta(\varepsilon_{\bm k}-\varepsilon_{\bm k'})  \left(1-\frac{v_{\bm{k}'}^{(i)}}{v_{\bm{k}}^{(i)}}\right)  \nonumber \\
&=& \frac{2\pi n_{\rm{imp}}}{\hbar}  \frac{g v k}{2 \pi \hbar v_x v_y} \bar{V}^2_{\rm agp},
\end{eqnarray}
where $\bar{V}^{2}_{\rm agp} \equiv \int_0^{2\pi}{\frac{d\phi'}{(2\pi)}}|V_0|^2 F(\phi') \left(1-\frac{v_{\bm{k}'}^{(i)}}{v_{\bm{k}}^{(i)}}\right)=\frac{|V_0|^2}{4}$ is the angle-averaged square of the impurity potential for the anisotropic graphene. Note that $\tau^{(i)}$ for the anisotropic Dirac semimetal is also independent of the direction $i$.

Therefore, the conductivity of the anisotropic Dirac semimetal is given by
\begin{eqnarray}
\sigma_{ij}&=&g e^2 \int{\frac{d^2\bm{k}}{(2\pi)^2}} \delta(\varepsilon_{\rm{F}}-\varepsilon(\bm{k}))v^{(i)} v^{(j)} \tau ^{(j)} \\ \nonumber 
&=& \frac{g e^2 v k}{2 \pi \hbar v_x v_y}\tau_{\rm{F}} \int_0^{2\pi}{\frac{d\phi}{2\pi}} v_{\rm{F}}^{(i)} v_{\rm{F}}^{(j)} .
\end{eqnarray}
When the electric field and the current density are along the $x$-direction, the conductivity $\sigma_{xx}$ becomes 
\begin{eqnarray}
&&\sigma_{xx} = \frac{g e^2 v k}{2 \pi \hbar v_x v_y} \tau_{\rm{F}} \int_0^{2\pi}{\frac{d\phi}{2\pi}} \left[v_{\rm{F}}^{(x)}\right]^2 \nonumber \\
&&= \frac{g e^2 v k}{2 \pi \hbar v_x v_y}  \frac{\hbar} {2\pi n_{\rm{imp}}} \frac{2 \pi \hbar v_x v_y}{v k \bar{V}^2_{\rm agp}} v_x^2 \int_0^{2\pi}{\frac{d\phi}{2\pi}} \cos^2 \phi \nonumber \\
&&=\frac{g e^2 \hbar v^2}{4\pi n_{\rm{imp}}\bar{V}^2_{\rm agp}} \frac{v_x^2}{v^2}.
\end{eqnarray}
Similarly, when the electric field and the current density are along the $y$-direction, the conductivity $\sigma_{yy}$ becomes 
\begin{eqnarray}
\sigma_{yy} &=&  \frac{g e^2 v k}{2 \pi \hbar v_x v_y} \tau_{\rm{F}} \int_0^{2\pi}{\frac{d\phi}{2\pi}} \left[v_{\rm{F}}^{(y)}\right]^2 \nonumber \\ 
&=& \frac{g e^2 \hbar v^2}{4\pi n_{\rm{imp}}\bar{V}^2_{\rm agp}} \frac{v_y^2}{v^2}.
\end{eqnarray}
Therefore, the dc conductivities for the anisotropic case are modified as
\begin{subequations}
\begin{eqnarray}
\sigma_{xx} &=& \sigma_{\rm{iso}} \frac{v_x^2}{v^2}, \\ 
\sigma_{yy} &=& \sigma_{\rm{iso}} \frac{v_y^2}{v^2}. 
\end{eqnarray}
\end{subequations}
Thus, for short-range impurities or charged-impurities in the strong screening limit, the Fermi energy dependence of the dc conductivities for the anisotropic graphene follows that of the isotropic graphene given by Eqs.~(\ref{eq:app_density_dependence_short_Dirac}) and (\ref{eq:app_density_dependence_charged_strong_screening_Dirac}).

Near the van Hove singularities, where the energy dispersion can be expanded as $\varepsilon(\bm{k})/\varepsilon_0 \approx |\Delta|-\tilde{k}_x^2+\frac{c^2 \tilde{k}_y^2}{2|\Delta|}$, the DOS diverges logarithmically \cite{Marder2010}, dominating the overall power-law behavior of the conductivity. Therefore, for short-range impurities, the conductivity becomes 
\begin{subequations}\label{eq:app_density_dependence_short_vanhove}
\begin{eqnarray}
\sigma_{xx}&\sim& [-\log(|\Delta|-\varepsilon_{\rm F})]^{-1}, \\
\sigma_{yy}&\sim& [-\log(|\Delta|-\varepsilon_{\rm F})]^{-1}.
\end{eqnarray}
\end{subequations}
For the charged impurities near the van Hove singularities, the conductivity becomes
\begin{subequations}\label{eq:app_density_dependence_charged_strong_screening_vanhove}
\begin{eqnarray}
\sigma_{xx}&\sim& [\log(|\Delta|-\varepsilon_{\rm F})]^{2}, \\
\sigma_{yy}&\sim& [\log(|\Delta|-\varepsilon_{\rm F})]^{2}.
\end{eqnarray}
\end{subequations}

Note that, as the Fermi energy or the carrier density increases, the power-law dependence of the dc conductivity follows that of the semi-Dirac transition point, as in the gapped insulator case. 

\section{Temperature dependence of chemical potential and Thomas--Fermi wave vector in black phosphorus}
\label{sec:chemical_potential_screening_wavevector}

In this section, we derive the temperature-dependent chemical potential and Thomas--Fermi wave vector of few-layer BP. 
When the temperature is finite, the chemical potential $\mu$ deviates from the Fermi energy $\varepsilon_{\rm F}$ due to the broadening of the Fermi distribution function $f^{(0)}(\varepsilon,\mu)=\left[e^{\beta (\varepsilon-\mu)}+ 1\right]^{-1}$ where $\beta={1\over k_{\rm B}T}$. As the charge carrier density $n$ does not vary under the temperature change, we have
\begin{eqnarray}
n&=&\int_{-\infty}^{\infty}d\varepsilon D(\varepsilon)f^{(0)}(\varepsilon ,\mu) \nonumber \\  
&=&\int_{0}^{\infty}d\varepsilon D(\varepsilon)\left[f^{(0)}(\varepsilon ,\mu)+f^{(0)}(-\varepsilon ,\mu)\right] \nonumber \\
&\equiv& \int_{-\infty}^{\varepsilon_{\rm F}}d\varepsilon D(\varepsilon).
\end{eqnarray}
Thus, the carrier density measured from the charge neutral point, $\Delta n\equiv \left.n\right|_{\mu}-\left.n\right|_{\mu=0}$, is given by
\begin{eqnarray}\label{eq:DensityInvar_DoubleBand}
\Delta n &=& \int_{0}^{\infty}d\varepsilon D(\varepsilon)\left[f^{(0)}(\varepsilon ,\mu)-f^{(0)}(\varepsilon ,-\mu)\right] \nonumber  \\
&\equiv& \int_{0}^{\varepsilon_{\rm F}}d\varepsilon D(\varepsilon),
\end{eqnarray}
where the first and second lines represent the carrier density evaluated at the finite and zero temperatures, respectively. Here, we used $f(-\varepsilon,\mu)=1-f(\varepsilon,-\mu)$. By solving this equality in terms of $\mu$, we can calculate the chemical potential of the system for a given temperature $T$. See the Supplemental Material in \cite{Park2017} for the simplified cases.

Subsequently, consider the temperature-dependent Thomas--Fermi wave vector $q_{\rm TF}(T)$. Note that, in 3D, $q_{\rm TF}(0)={2\pi e^2\over \kappa} D(\varepsilon_{\rm F})$ and at finite $T$, $q_{\rm TF}(T)={2\pi e^2\over \kappa} {\partial n \over \partial \mu}$. Thus, we have
\begin{eqnarray}\label{eq:q_TF_DoubleBand}
&& \frac{q_{\rm{TF}}(T) }{q_{\rm{TF}}(0) }  = \frac{ \beta }{2D(\varepsilon_{\rm F})} \int_0^{\infty}{d\varepsilon} D(\varepsilon) \\ \nonumber 
&&\times \left[ \frac{ 1 }{1+\cosh \beta(\varepsilon - \mu)}+\frac{ 1 }{1+\cosh  \beta(\varepsilon + \mu)} \right] .
\end{eqnarray}
For a given $T$, the chemical potential is calculated using the density invariance in Eq.~(\ref{eq:DensityInvar_DoubleBand}), and subsequently, $q_{\rm{TF}}(T)$ is obtained from the above relation.

When the DOS is given by a simple power law with respect to energy, we can analytically obtain the temperature dependence of the chemical potential and Thomas--Fermi wave vector, and their asymptotic behaviors at low and high temperatures.

Consider a gapless electron--hole system with a DOS given by $D(\varepsilon) = C_\alpha |\varepsilon|^{\alpha-1}\varTheta(\varepsilon)$, where $C_\alpha$ is a constant and $\varTheta(\varepsilon)$ is a step function. 
Using the results from the Supplemental Materials in Ref.~\cite{Park2017}, we can obtain
\begin{eqnarray}\label{eq:mu_temperature_correction}
{\frac{\mu}{\varepsilon_{\rm F}}} &=&
\begin{cases}
1- \frac{\pi^2}{12} \left(\frac{T}{T_{\rm F}} \right)^2 & (T\ll T_{\rm F}), \\ 
\frac{1}{2\eta(\alpha-1)\Gamma(\alpha+1)}\left(\frac{T_{\rm F}}{T}\right)^{\alpha-1} & (T\gg T_{\rm F}),
\end{cases}
\end{eqnarray}
where $T_{\rm F}=\varepsilon_{\rm F}/k_{\rm B}$ is the Fermi temperature, $\eta$ is the Dirichlet eta function, and $\Gamma$ is the gamma function \cite{Arfken2012}.
For the temperature-dependent Thomas--Fermi wave vector $q_{\rm TF}(T)$, we obtain
\begin{eqnarray}\label{eq:mu_qTF_correction}
{\frac{q_{\rm{TF}}(T) }{q_{\rm{TF}}(0) }} &=&
\begin{cases}
1-{\pi^2\over 6}(\alpha-1)\left(\frac{T}{T_{\rm F}} \right)^2 &\!\! (T\ll T_{\rm F}), \\ 
2\eta(\alpha-1) \Gamma(\alpha)\left(\frac{T}{T_{\rm F}} \right)^{\alpha-1} &\!\! (T\gg T_{\rm F}),
\end{cases}
\end{eqnarray}

For few-layer BP at the semi-Dirac transition point, the DOS is given by $D(\varepsilon) \propto \varepsilon^{1 \over 2}$; thus, $\alpha =\frac{3}{2}$. 
Thus, we have
\begin{eqnarray}\label{eq:mu_bp_temperature_correction}
{\frac{\mu}{\varepsilon_{\rm F}}} &=&
\begin{cases}
1- \frac{\pi^2}{12} \left(\frac{T}{T_{\rm F}} \right)^2 & (T\ll T_{\rm F}), \\ {1\over 2\eta\left({1\over 2}\right)\Gamma\left({5\over 2}\right)}\left(\frac{T}{T_{\rm F}}\right)^{{1\over 2}} & (T\gg T_{\rm F}),
\end{cases}
\end{eqnarray}
and
\begin{eqnarray}\label{eq:q_TF_BP_temperature_correction}
{\frac{q_{\rm{TF}}(T) }{q_{\rm{TF}}(0) }} &= &
\begin{cases}
1- \frac{\pi^2}{12} \left(\frac{T}{T_{\rm F}} \right)^2  & (T\ll T_{\rm F}), \\
2\eta\left({1\over 2}\right)\Gamma\left({3\over 2}\right)\left(\frac{T}{T_{\rm F}}\right)^{1\over 2} & (T\gg T_{\rm F}),
\end{cases}
\end{eqnarray}
where $q_{\rm TF}(0)=q_{\rm TF}$ is given by Eq.~(\ref{eq:q_TF_mWSM}).

\begin{figure}[hbt]
\includegraphics[width=0.8\linewidth]{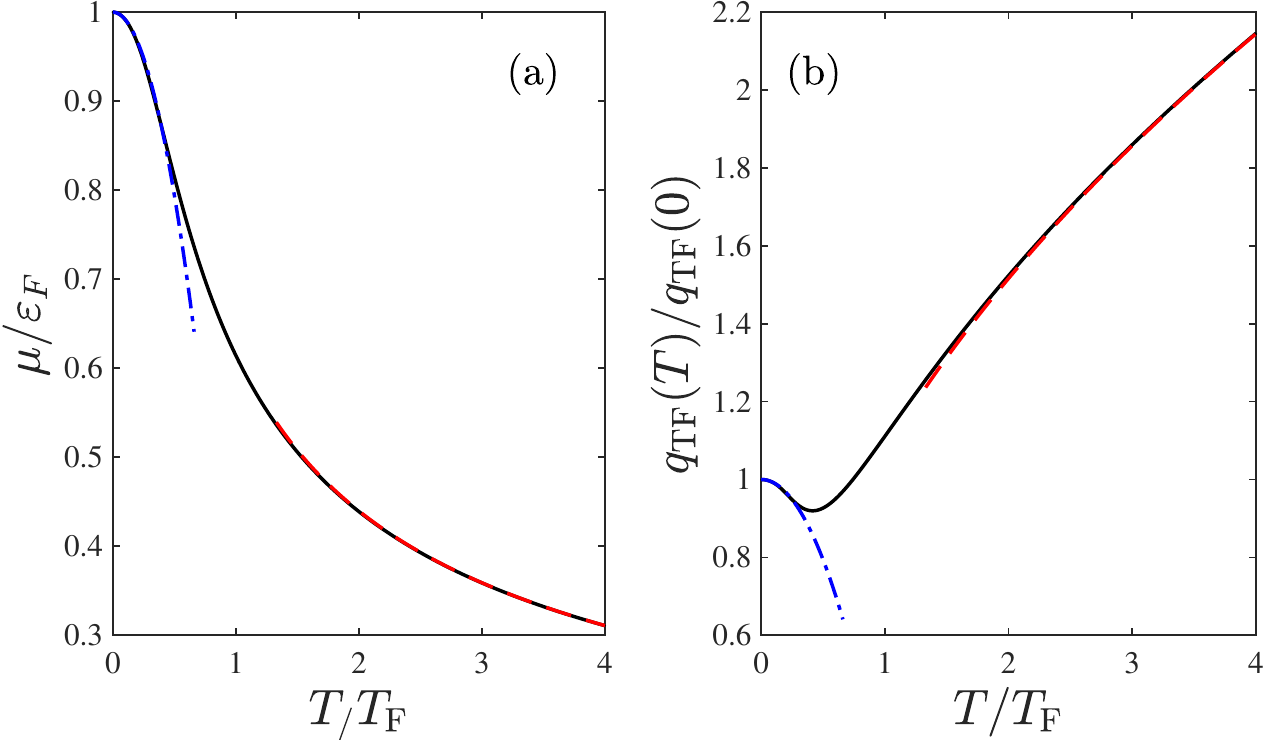}
\caption{
Calculated temperature dependence of (a) chemical potential and (b) Thomas--Fermi wave vector for the semi-Dirac transition point ($\Delta=0$). Here, the black solid lines represent the numerical result, and the red/blue dashed lines represent the high-/low-temperature asymptotic forms in Eqs.~(\ref{eq:mu_bp_temperature_correction}) and (\ref{eq:q_TF_BP_temperature_correction}). If the chemical potential and temperature are normalized by $\varepsilon_{\rm F}$ and $T_{\rm F}$, respectively, the result is independent of $\varepsilon_{\rm F}$ at the semi-Dirac transition point.
}
\vspace{10pt}%
\label{fig:mu_qtf_semidirac}
\end{figure}

\begin{figure}[hbt]
\includegraphics[width=\linewidth]{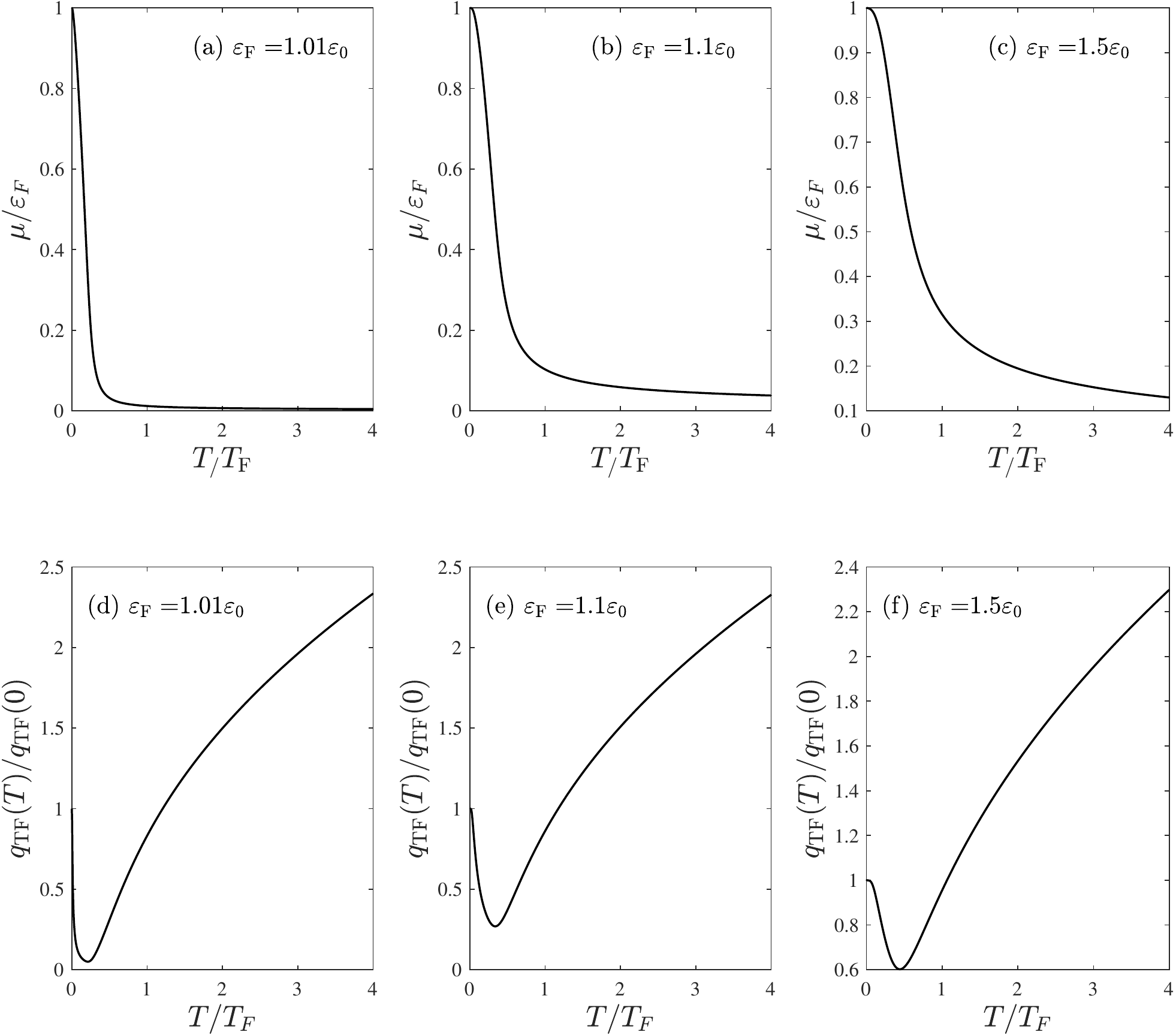}
\caption{
Calculated temperature dependence of (a)-(c) chemical potential and (d)-(f) Thomas--Fermi wave vector for the gapped insulator phase with $\Delta=1$ at (a), (d) $\varepsilon_{\rm F}=1.01\varepsilon_0$, (b), (e) $\varepsilon_{\rm F}=1.1\varepsilon_0$, and (c), (f) $\varepsilon_{\rm F}=1.5\varepsilon_0$.
}
\vspace{25pt}
\label{fig:mu_qtf_semicon}
\end{figure}

\begin{figure}[bt]
\vspace{-20pt}
\includegraphics[width=\linewidth]{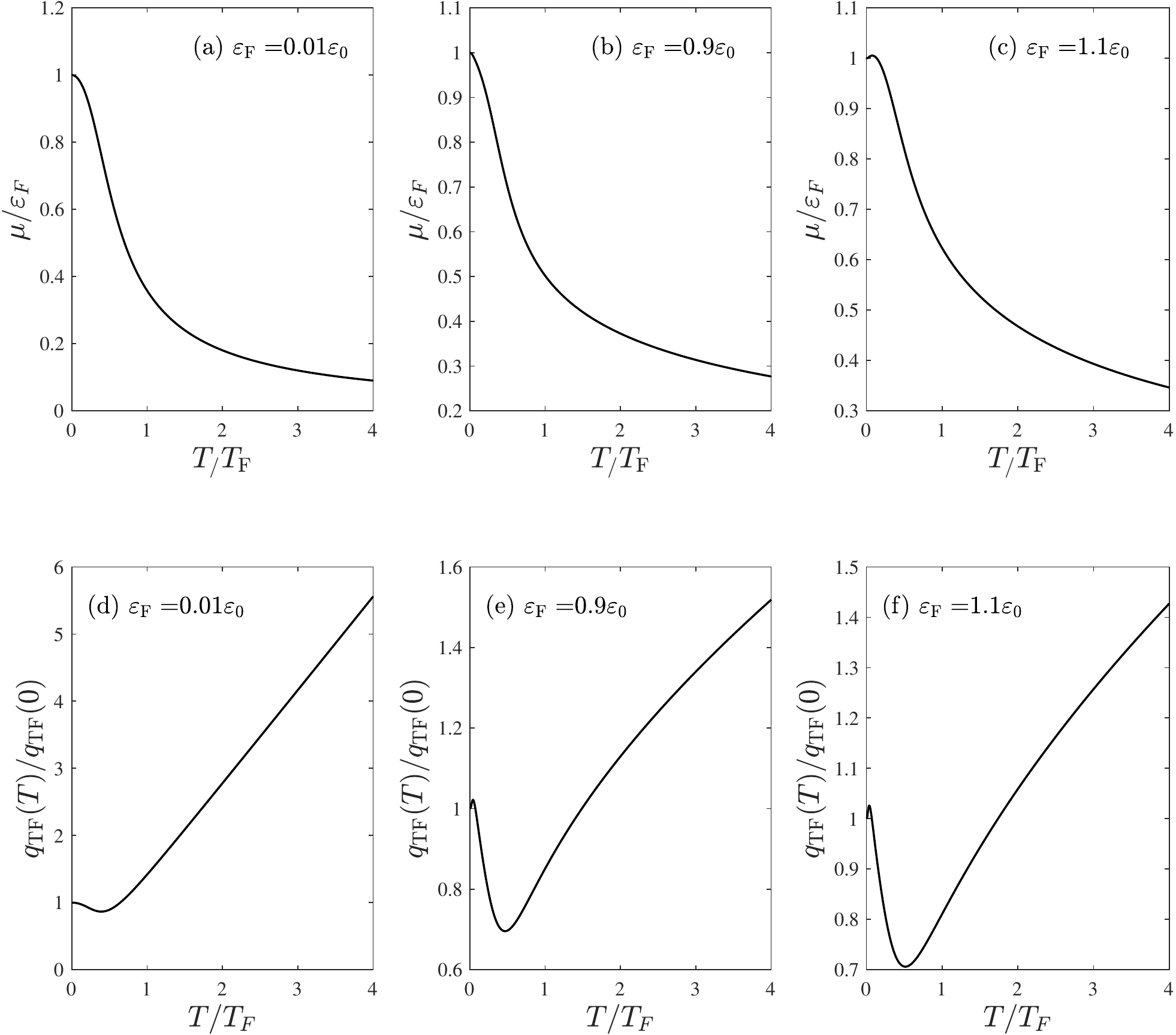}
\caption{
Calculated temperature dependence of (a)-(c) chemical potential and (d)-(f) Thomas--Fermi wave vector for the Dirac semimetal phase with $\Delta=-1$ at (a), (d) $\varepsilon_{\rm F}=0.01\varepsilon_0$, (b), (e) $\varepsilon_{\rm F}=0.9\varepsilon_0$, and (c), (f) $\varepsilon_{\rm F}=1.1\varepsilon_0$.
}
\label{fig:mu_qtf_dirac}
\end{figure}

Figure~\ref{fig:mu_qtf_semidirac}, Figure~\ref{fig:mu_qtf_semicon}, and Figure~\ref{fig:mu_qtf_dirac} show the calculated temperature dependence of the chemical potential $\mu(T)$ and Thomas--Fermi wave vector $q_{\rm TF}(T)$ in various phases of BP using Eqs.~(\ref{eq:DensityInvar_DoubleBand}) and (\ref{eq:q_TF_DoubleBand}), respectively.

\section{Temperature dependence of dc conductivity at the semi-Dirac transition point}
\label{sec:app_temperature_dep_semidirac}

Using Eq.~(\ref{eq:conductivity_tensor}) in the main text, we can generalize the conductivity tensor at zero temperature to that at finite temperature. 
For $f^{(0)}(\varepsilon)=\left[z^{-1}e^{\beta \varepsilon}+ 1\right]^{-1}$, where $z=e^{\mu}$ is the fugacity, $S^{(0)}(\varepsilon)=-{\partial f^{(0)}(\varepsilon)\over\partial\varepsilon}=\beta f^{(0)}(\varepsilon) \left[1-f^{(0)}(\varepsilon)\right]={\beta z^{-1} e^{\beta \varepsilon} \over (z^{-1} e^{\beta\varepsilon}+ 1)^2}$. 
Thus, the conductivity at finite temperature is given by

\begin{widetext}
\begin{eqnarray}
\label{eq:conductivity_finite_temperature}
\sigma_{ij}(T)&=&g e^2 \sum_{\alpha} \int {d^2 k\over (2\pi)^2} \left(-{\partial f^{(0)}(\varepsilon_{\bm{k}})\over\partial\varepsilon}\right) v_{\bm{k}\alpha}^{(i)} v_{\bm{k\alpha}}^{(j)}\tau_{\bm{k}\alpha}^{(j)} \nonumber \\
&=&g e^2 \sum_{\alpha}\int_0^{\infty}dr  \int_r'd\phi
\frac{ k_0^2r}{2c\sqrt{r \cos\phi-\Delta}} 
{\beta z^{-1} e^{\beta \varepsilon_0 r} \over (z^{-1} e^{\beta\varepsilon_0 r}+ 1)^2} v_{\bm{k}\alpha}^{(i)} v_{\bm{k}\alpha}^{(j)}\tau_{\bm{k}\alpha}^{(j)} \nonumber \\
&=&\sigma_0\sum_{\alpha} \int_0^{\infty}dr \int_r' {d\phi\over 2\pi} \frac{r}{2 \sqrt{r \cos\phi-\Delta}}  {\beta\varepsilon_0 z^{-1} e^{\beta \varepsilon_0 r} \over (z^{-1} e^{\beta\varepsilon_0 r}+ 1)^2}
\tilde{v}_{\bm{k}\alpha}^{(i)} \tilde{v}_{\bm{k}\alpha}^{(j)} \tilde{\tau}_{\bm{k}\alpha}^{(j)}.
\end{eqnarray}
Thus, from Eq.~(\ref{eq:band_velocity}), we have
\begin{subequations}
\label{eq:conductivity_finite_temperature_component}
\begin{eqnarray}
\sigma_{xx}(T)&=&2 \sigma_0 \sum_{\alpha} \int_0^{\infty} dr\, {\beta\varepsilon_0 z^{-1} e^{\beta \varepsilon_0 r} \over (z^{-1} e^{\beta\varepsilon_0 r}+ 1)^2}\int_r'\frac{d\phi}{2\pi} r \cos^2\phi \sqrt{r \cos\phi-\Delta} \tilde{\tau_{\alpha}}^{(x)}(\phi), \\
\sigma_{yy}(T)&=& c^2 \sigma_0 \sum_{\alpha}  \int_0^{\infty} dr\, {\beta\varepsilon_0 z^{-1} e^{\beta \varepsilon_0 r} \over (z^{-1} e^{\beta\varepsilon_0 r}+ 1)^2} \int_r'\frac{d\phi}{2\pi} \frac{r \sin^2\phi}{2\sqrt{r \cos\phi-\Delta}} \tilde{\tau}_{\alpha}^{(y)}(\phi).
\end{eqnarray}
\end{subequations}
\end{widetext}

To derive the asymptotic behaviors of $\sigma_{ii}(T)$ at low and high temperatures, assume that the relaxation time can be decomposed into energy- and temperature-dependent parts as $\tau^{(i)}(\varepsilon,T)=\tau^{(i)}(\varepsilon) g^{(i)}\left(\frac{T}{T_{\rm F}}\right)$ where $g^{(i)}\left(\frac{T}{T_{\rm F}}\right)$ is the energy-independent correction term from the temperature-dependent screening effect with $g^{(i)}(0)=1$. For short-range impurities, $g^{(i)}\left(\frac{T}{T_{\rm F}}\right)=1$. For charged Coulomb impurities, we expect $g\left(\frac{T}{T_{\rm F}}\right) \approx 1-A^{(i)} \left(\frac{T}{T_{\rm F}}\right)^2$ at low temperatures, whereas at high temperatures, the energy averaging typically dominates over the screening contribution and we can assume $g\left(\frac{T}{T_{\rm F}}\right)\approx 1$. 
Suppose the following power-law dependence: $D(\varepsilon)\sim \varepsilon^{\alpha-1}$, $v^{(i)}(\varepsilon) \sim \varepsilon^{\nu}$, and $\tau^{(i)}(\varepsilon)\sim \varepsilon^{\gamma}$. Subsequently, by rewriting Eq.~(\ref{eq:conductivity_finite_temperature}) as an energy-integral form, we obtain the asymptotic power-law behavior of the temperature-dependent conductivity at low and high temperatures as
\begin{eqnarray}\label{eq:asymptotic_temp}
{\sigma_{ii}(T)\over \sigma_{ii}(0)} &=&
\begin{cases}
1+\left[\frac{\pi^2}{6}(\delta-\alpha)\delta-A^{(i)}\right] \left(\frac{T}{T_{\rm F}}\right)^2  & (T\ll T_{\rm F}), \nonumber \\ 
\Gamma(\delta+1) \eta(\delta)\left(\frac{T}{T_{\rm F}}\right)^{\delta}  & (T\gg T_{\rm F}),
\end{cases} \nonumber \\
\end{eqnarray}
where $\delta=\alpha-1+2\nu+\gamma$. (See the Supplemental Material of Ref.~\cite{Park2017} for the detailed derivation of the power-law analysis of the temperature-dependent dc conductivity.) For the semi-Dirac transition point ($\Delta=0$), $\alpha=\frac{3}{2}$. 

For short-range impurities, $g\left(\frac{T}{T_{\rm F}}\right)=1$ and from the energy dependence of the relaxation time, $\gamma=-\frac{1}{2}$. Thus, from Eq.~(\ref{eq:conductivity_finite_temperature_component}), the asymptotic behavior is given by
\begin{subequations}\label{eq:asymptotic_temp_semidirac_short_app}
\begin{eqnarray}
{\sigma_{xx}(T)\over \sigma_{xx}(0)} &=&
\begin{cases}
1-\frac{\pi^2}{12} \left(\frac{T}{T_{\rm F}}\right)^2  & (T\ll T_{\rm F}), \\ 
\log2 \left(\frac{T}{T_{\rm F}}\right)  & (T\gg T_{\rm F}),
\end{cases} \\
{\sigma_{yy}(T)\over \sigma_{yy}(0)} &=&
\begin{cases}
1-e^{-T_{\rm F}/T}  &  \!\! (T\ll T_{\rm F}), \\ 
\frac{1}{2} + \frac{1}{8\eta\left({1\over 2}\right)\Gamma\left({5\over 2}\right)} \left(\frac{T}{T_{\rm F}}\right)^{-{3\over 2}} & \!\! (T\gg T_{\rm F}).
\end{cases} 
\end{eqnarray}
\end{subequations}
Here, the extra terms in $\sigma_{yy}(T)/\sigma_{yy}(0)$ were obtained through the next-order expansion of the temperature corrections.

For charged impurities in the strong screening limit, $A^{(i)}=\frac{\pi^2}{6}$, which is two times the low-temperature coefficient $\frac{\pi^2}{12}$ in Eq.~(\ref{eq:q_TF_BP_temperature_correction}), and $\gamma=\frac{1}{2}$. Thus, we obtain 
\begin{subequations}
\begin{eqnarray}\label{eq:asymptotic_temp_semidirac_long_app}
{\sigma_{xx}(T)\over \sigma_{xx}(0)} &=&
\begin{cases}
1 & (T\ll T_{\rm F}), \\ 
\frac{\pi^2}{6} \left(\frac{T}{T_{\rm F}}\right)^2  & (T\gg T_{\rm F}),
\end{cases} \\
{\sigma_{yy}(T)\over \sigma_{yy}(0)} &=&
\begin{cases}
1-\frac{\pi^2}{4} \left(\frac{T}{T_{\rm F}}\right)^2  & (T\ll T_{\rm F}), \\ 
\log2 \left(\frac{T}{T_{\rm F}}\right) & (T\gg T_{\rm F}).
\end{cases} 
\end{eqnarray}
\end{subequations}
As the screening strength decreases, the low-temperature coefficient in Eq.~(\ref{eq:asymptotic_temp}) increases, because the screening coefficient $A^{(i)}$ decreases whereas the other part remains positive.

\section{Temperature dependence of dc conductivity in the low-density approximate models for the insulator phase and Dirac semimetal phase}
\label{sec:app_temperature_dep_effmodel}

In this section, we present the temperature dependence of the chemical potential, Thomas--Fermi wave vector, and conductivity of the low-density approximate models for the insulator phase and Dirac semimetal phase, which are the gapped 2DEG and graphene, respectively.

\subsection{Insulator phase}

We introduce the gapped 2DEG model system with the energy dispersion given by $\varepsilon({\bm k})= \pm \varepsilon_0\left[\Delta+(k/k_0)^2\right]$ with $\Delta>0$, to account for the thermal excitation behavior involving the band gap between the valence and conduction bands, similar to the insulator phase. Note that the effects of the difference between the effective mass of each direction are canceled out by zero-temperature normalization.

\begin{figure}[htb]
\includegraphics[width=\linewidth]{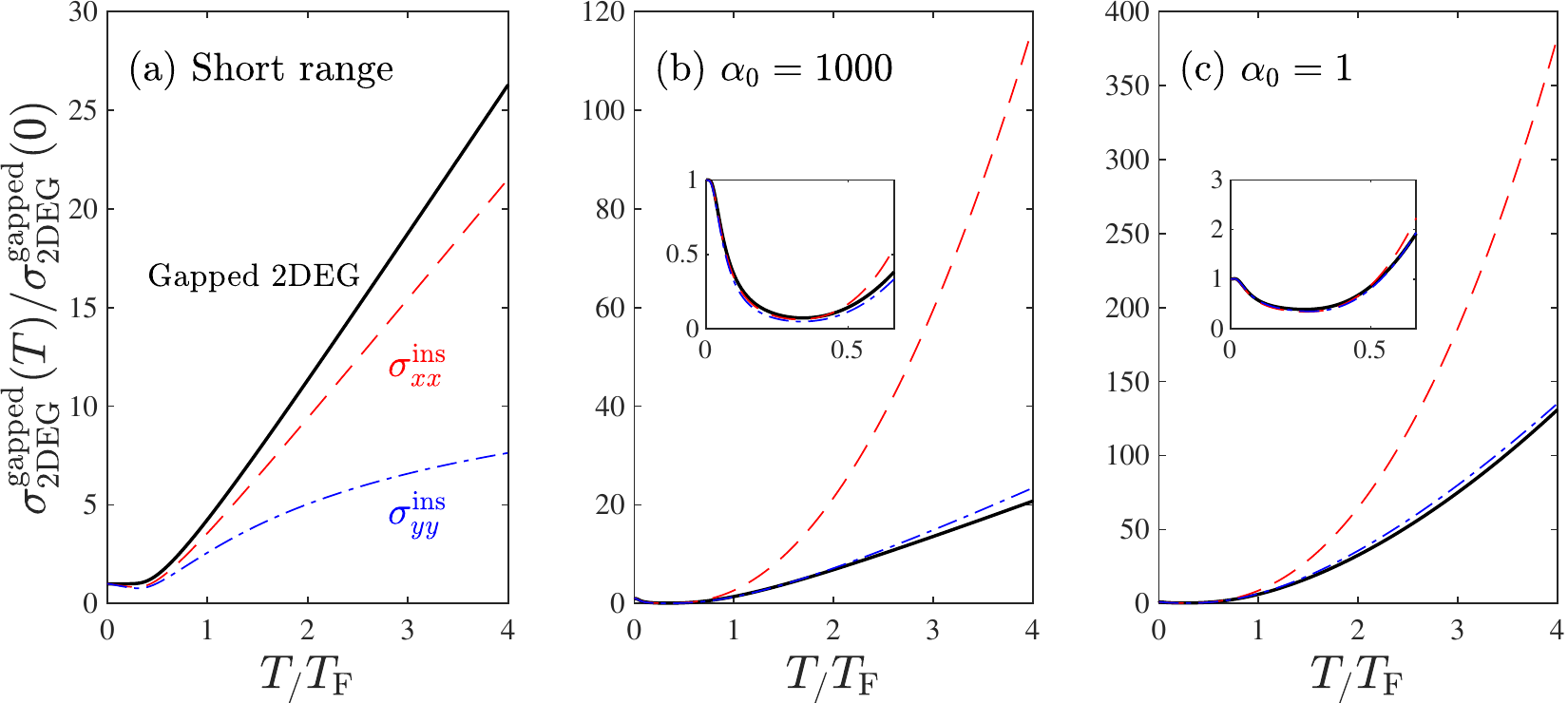}
\caption{
Calculated dc conductivities as a function of the temperature for the gapped 2DEG system in the low-density limit with $\Delta=1$ for (a) short-range impurities, (b) charged impurities with $\alpha_0=1000$, and (c) charged impurities with $\alpha_0=1$. Here, $\varepsilon_{\rm F}=1.1 \varepsilon_0$ is used for the calculation. The red dashed lines and blue dashed-dotted lines represent the conductivity of the insulator phase (with the same Fermi energy) $\sigma_{xx}^{\rm ins}$ and $\sigma_{yy}^{\rm ins}$, respectively. 
}
\label{fig:temp_2DEG_110}
\end{figure}

Figure~\ref{fig:temp_2DEG_110} shows the calculated dc conductivities as a function of the temperature for the gapped 2DEG system in the low-density limit with $\varepsilon_{\rm F}=1.1 \varepsilon_0$ along with the result of the insulator phase with the same Fermi energy (see also Fig.~\ref{fig:temp_semicon_110} in the main text).
At low temperatures, the calculated results of temperature-dependent conductivity in the insulator phase show a similar behavior as that of the low-density approximate model. 
However, as the temperature increases, the discrepancy between the two results increases, and in the high-temperature limit, the conductivity becomes similar to that of the semi-Dirac transition point.

\subsection{Dirac semimetal Phase}

For graphene (which is an approximate model for the Dirac semimetal phase in the low-density limit) from Eqs.~(\ref{eq:mu_temperature_correction}) and (\ref{eq:mu_qTF_correction}) with $\alpha=2$, the low-- and high--temperature asymptotic behaviors for chemical potential are given by
\begin{eqnarray}\label{eq:asymptotic_mu_graphene}
{\frac{\mu}{\varepsilon_{\rm F}}} &=&
\begin{cases}
1- \frac{\pi^2}{6} \left(\frac{T}{T_{\rm F}} \right)^2 & (T\ll T_{\rm F}), \\ 
\frac{1}{4 \log 2}\left(\frac{T}{T_{\rm F}}\right)^{-1} & (T\gg T_{\rm F}),
\end{cases}
\end{eqnarray}
whereas those for the Thomas--Fermi wave vector are given by
\begin{eqnarray}\label{eq:asymptotic_qtf_graphene}
{\frac{q_{\rm{TF}}(T) }{q_{\rm{TF}}(0) }} &= &
\begin{cases}
1- \frac{\pi^2}{6} \left(\frac{T}{T_{\rm F}} \right)^2  & (T\ll T_{\rm F}), \\
2\log 2\left(\frac{T}{T_{\rm F}}\right)^{} & (T\gg T_{\rm F}).
\end{cases}
\end{eqnarray}
As shown in Figs.~\ref{fig:mu_qtf_dirac}(a) and \ref{fig:mu_qtf_dirac}(d), the result of the low-density approximate model and the numerically calculated result of the Dirac semimetal phase in the low-density limit are consistent with each other. 
For short-range impurities, the asymptotic form of the temperature-dependent conductivity becomes [Eq.~(\ref{eq:asymptotic_temp_semidirac_short_app}) with $\gamma=0$] 
\begin{eqnarray}\label{eq:asymptotic_temp_dirac_short_app}
{\sigma_{\rm gp}(T)\over \sigma_{\rm gp}(0)} &=&
\begin{cases}
1-e^{-T_{\rm F}/T}  & (T\ll T_{\rm F}), \\ 
\frac{1}{2}+ \frac{1}{16 \log 2} \left(\frac{T}{T_{\rm F}}\right)^{-2} & (T\gg T_{\rm F}),
\end{cases}
\end{eqnarray}
whereas for charged impurities in the strong screening limit, [Eq.~(\ref{eq:asymptotic_temp_semidirac_short_app}) with $\gamma=2$]
\begin{eqnarray}\label{eq:asymptotic_temp_dirac_long_app}
{\sigma_{\rm gp}(T)\over \sigma_{\rm gp}(0)} &=&
\begin{cases}
1-\frac{\pi^2}{3} \left(\frac{T}{T_{\rm F}}\right)^{2} & (T\ll T_{\rm F}), \\ 
\frac{\pi^2}{6} \left(\frac{T}{T_{\rm F}}\right)^{2} & (T\gg T_{\rm F}).
\end{cases}
\end{eqnarray}
\\




\end{document}